\documentclass[reprint,amsmath,amssymb,aps,showkeys]{revtex4-2}
\usepackage{graphicx}
\usepackage{subfigure}
\usepackage{dcolumn}
\usepackage{bm}
\usepackage{siunitx}
\usepackage{multirow}
\usepackage{booktabs}
\usepackage{float}
\usepackage{hyperref}
\usepackage{cleveref}
\hypersetup{colorlinks = true, linkcolor = black, urlcolor = blue, citecolor = blue}

\begin{document}

\title{Triply charmed baryons mass decomposition from lattice QCD}

\author{Jin-Bo Li} 

\affiliation{  Department
of Physics, Hunan Normal University,  Changsha 410081, China }

\author{Long-Cheng Gui} \email{guilongcheng@hunnu.edu.cn}

\affiliation{  Department
of Physics, Hunan Normal University,  Changsha 410081, China }

\affiliation{ Synergetic Innovation
Center for Quantum Effects and Applications (SICQEA), Changsha 410081, China}

\affiliation{  Key Laboratory of
Low-Dimensional Quantum Structures and Quantum Control of Ministry
of Education, Changsha 410081, China}

\author{Wei Sun } %
\email{sunwei@ihep.ac.cn}
\affiliation{ Institute of High Energy Physics, Chinese Academy of Sciences, Beijing 100049, P.R. China }

\author{Jian Liang } \email{jianliang@scnu.edu.cn} %
\affiliation{ Key Laboratory of Atomic and Subatomic Structure and Quantum Control (MOE), Guangdong Basic Research Center of Excellence for Structure and Fundamental Interactions of Matter, Institute of Quantum Matter, South China Normal University, Guangzhou 510006, China
}

\affiliation{ Guangdong-Hong Kong Joint Laboratory of Quantum Matter, Guangdong Provincial Key Laboratory of Nuclear Science, Southern Nuclear Science Computing Center, South China Normal University, Guangzhou 51006, China}

\author{Wen Qin } \email{qinwen@hunnu.edu.cn}
\affiliation{  Department
of Physics, Hunan Normal University,  Changsha 410081, China }

\affiliation{ Synergetic Innovation
Center for Quantum Effects and Applications (SICQEA), Changsha 410081, China}

\affiliation{  Key Laboratory of
Low-Dimensional Quantum Structures and Quantum Control of Ministry
of Education, Changsha 410081, China}

\begin{abstract}

We present the first calculation of the connected scalar matrix element
and the momentum fraction of charm quark within the $\frac{3}{2}^{+}$
and $\frac{3}{2}^{-}$ triply charmed baryons on lattice QCD. 
The results are based on overlap valence fermions on two ensembles of $N_f=2+1$ domain wall fermion configurations with two lattice spacings. The corresponding sea quark pion masses are $300$ MeV and $278$ MeV.
The separated
contributions to the triply charmed baryon mass are derived through
the decomposition of the QCD energy-momentum tensor. The contribution
of the connected charm quark matrix element to the triply charmed
baryon is about 3/2 times that of the charmonium. And it contributes
almost 70\% of the total mass. The mass splitting of $\frac{3}{2}^{+}$ 
and $\frac{3}{2}^{-}$ triply charmed baryons is mainly from $\langle H_{E}\rangle$
of the QCD energy-momentum tensor. A mass decomposition from quark
model is also studied as a comparison.

\end{abstract}


\maketitle

\section{Introduction}{\label{introduction}}

Since the discovery of the $J/\psi$ meson in 1974~\citep{E598:1974sol,SLAC-SP-017:1974ind},
charm physics has garnered significant attention, leading to the identification
of numerous charmed hadrons. Recently, the Large Hadron Collider (LHC)
has made notable advancements in the study of charmed baryons, including
the spectroscopy of excited hadrons such as $\Lambda_{c}^{+}$~\citep{LHCb:2017jym},
$\Omega_{c}(X)$~\citep{LHCb:2017uwr}, and $\Xi_{c}(X)$~\citep{LHCb:2020iby}.
Notably, the doubly charmed baryon $\Xi_{cc}^{++}$ was established
by LHCb in 2017~\citep{LHCb:2017iph} and has since been confirmed
by subsequent studies~\citep{LHCb:2018pcs,LHCb:2022rpd}, following
the initial identification of the charmed baryon $\Lambda_{c}$ by
Mark-II in 1979~\citep{Abrams:1979iu}. However, the triply charmed
baryon has yet to be observed experimentally, although there are theoretical researches indicate that its discovery is promising at LHC~\cite{Chen:2011mb,Wang:2018utj}.

Despite the lack of experimental evidence, theoretical investigations
into the triply charmed baryon have been actively pursued using various
approaches, including the quark model~\citep{Silvestre-Brac:1996myf,Roberts:2007ni,Vijande:2015faa,Yang:2019lsg,Liu:2019vtx,Shah:2017jkr,Patel:2008mv}, QCD
sum rules~\citep{Zhang:2009re,Wang:2011ae,Aliev:2014lxa}, Faddeev equations~\citep{Sanchis-Alepuz:2011xjl,Radin:2014yna,Qin:2019hgk}, the di-quark
model~\citep{Thakkar:2016sog,Yin:2019bxe}, the variational method~\citep{Jia:2006gw,Flynn:2011gf},
the bag model~\citep{Bernotas:2008bu,Hasenfratz:1980ka}, Regge theory~\citep{Wei:2015gsa},
and the Bethe-Salpeter equation~\citep{Migura:2006ep}, etc. These studies
predict the mass of the ground state triply charmed baryon to be approximately
4.8 GeV. Numerous theoretical calculations have also been conducted
to determine the mass spectrum under both quenched \citep{Chiu:2005zc}
and unquenched lattice QCD \citep{Briceno:2012wt,PACS-CS:2013vie,Padmanath:2013zfa,Brown:2014ena,Can:2015exa,Alexandrou:2014sha,Bahtiyar:2020uuj,Alexandrou:2017xwd,Alexandrou:2012xk,Durr:2012dw,Chen:2017kxr}.
The predicted ground state mass from these lattice QCD calculations
is consistent with other theoretical predictions.

Beyond spectroscopy, decomposing the mass of a hadron within the QCD
framework offers deeper insight into its internal structure. This
approach provides valuable information about the distribution of mass
contributions from various components within the hadron. The mass
decomposition of the nucleon was first proposed in~\citep{Ji:1994av},
taking into account the dynamic interactions between quarks and gluons.
In this framework, hadron mass is divided into four components: quark energy, quark mass, gluon energy, and trace anomaly. Mass decomposition has been explored in various hadronic systems, yielding novel insights~\citep{Yang:2014xsa,Yang:2018nqn,Sun:2020pda}.
For instance, it has been found that the trace anomaly contributes
significantly in light hadron systems~\citep{Yang:2014xsa}, and the
quark mass matrix contributes less in hybrids compared to charmonium~\citep{Sun:2020pda}. In this work, we aim to study the mass decomposition
of the ground state triply charmed baryon using lattice QCD. We will
first calculate the hadron mass $M$, the valence charm quark mass
contribution $\langle H_{m}\rangle$, and the valence charm quark
momentum fraction $\langle x\rangle_{q}$ using lattice QCD. Subsequently,
the other components of the hadron mass can be determined through
the decomposition formula of the QCD energy-momentum tensor (EMT)
and the trace sum rule. 

There is also a mass decomposition in the non-relativistic quark model,
which separates the hadron mass into three parts: the mass and kinetic
energy of the constituent quarks, and the potential energy between
them. The hyper-fine splitting is primarily attributed to the spin-spin
interaction arising from the one-gluon exchange potential. Previous
results for the heavy meson mass decomposition from lattice QCD appear
to align with the constituent quark model's perspective~\citep{Yang:2014xsa}.
In this work, we will also study the mass decomposition in the constituent
quark model to give a comparison, and try to get some clues of the
correlation between the phenomenological model and QCD theory, deepen
our understanding of their intrinsic nature.
Given that the triply charmed baryon actives in the 
energy region that non-perturbative interactions cannot be ignored, a precise 
decomposition of the mass will provide valuable insights into comprehending 
the non-perturbative properties of QCD.

The remainder of our article is organized as follows: In Sec.~\ref{Formalism},
we provide a detailed introduction to the formula for hadron mass
decomposition within the QCD framework. We also discuss the relationship
between the hadron mass and two-point functions, as well as the hadron
matrix elements and three-point functions. In Sec.~\ref{Numerical details},
we present the specifics of our numerical simulation, including the
configuration information, effective mass, effective matrix element
calculations, and our fitting results. In Sec.~\ref{Discussion},
we analyze and discuss our results, including the comparison of the
constituent quark model. Finally, a brief summary is provided in Sec.~\ref{Summary}.

\section{Formalism} \label{Formalism}
\subsection{Mass decomposition from the QCD EMT}\label{Mass decomposition from the QCD EMT}

In this paper, we adopt the decomposition of QCD energy-momentum
tensor proposed by Ji~\citep{Ji:1994av}, which is also used in Ref.~\citep{Yang:2014xsa,Yang:2018nqn,Sun:2020pda}.
The energy-momentum tensor of QCD is written as

\begin{equation}
T^{\mu\nu}=\frac{1}{2}\bar{\psi}i\overleftrightarrow{D}^{(\mu}\gamma^{\nu)}\psi+\frac{1}{4}g^{\mu\nu}F^{2}-F^{\mu\alpha}F_{\alpha}^{\nu},\label{eq:EMT}
\end{equation}
where () symmetrizing all the indices, $\overleftrightarrow{D}^{\mu}=\overrightarrow{D}^{\mu}-\overleftarrow{D}^{\mu}$, $\overrightarrow{D}$
is the gauge-covariant derivative, $F^{\mu\nu}$ is the color field
strength tensor. The QCD Hamiltonian and the hadron mass could be
written in terms of the energy-momentum tensor 
\begin{equation}
H_{\mathrm{QCD}}=\int d^{3}xT^{00}(0,\mathbf{x}),\label{eq:Hamiltonian}
\end{equation}
\[
M=\frac{\langle H|H_{QCD}|H\rangle}{\langle H|H\rangle}\equiv\langle T^{00}\rangle,
\]
where the hadron state $|H\rangle$ is renormalized as $\langle H|H\rangle=2E(2\pi)^{3}\delta^{3}(0)$.

The hadron mass can be decomposed as 
\begin{equation}
M=\langle T^{00}\rangle=\langle H_{M}\rangle+\langle H_{E}^{(\mu)}\rangle+\langle H_{g}^{(\mu)}\rangle+\frac{1}{4}\langle H_{a}\rangle.\label{eq:decompositon}
\end{equation}
in the rest frame of the hadron state with 
\begin{align}
H_{E}^{(\mu)} & =\sum_{f}\int d^{3}x\bar{\psi}^{(f)}(\vec{D}\cdot\vec{\gamma})\psi^{(f)},\\
H_{M} & =\sum_{f}\int d^{3}x\bar{\psi}^{(f)}m_{f}\psi^{(f)},\\
H_{g}^{(\mu)} & =\int d^{3}x\frac{1}{2}(B^{2}-E^{2}),\\
H_{a} & =\int d^{3}x[\gamma_{m}\sum_{f}\bar{\psi}^{(f)}m_{f}\psi^{(f)}-\frac{\beta(g)}{g}(B^{2}+E^{2})],
\end{align}
where $\sum_{f}$ denotes the
summation of quark flavors, $\gamma_{m}$ is the quark mass anomalous
dimension, $\beta(g)$ is the $\beta$ function of QCD. $H_{E}$,
$H_{M}$, $H_{g}$ and $H_{a}$ denote the contributions from the quark
energy, quark condensate, the gluon field energy, and the joint contribution
of quantum anomalies from both gluon and quark in Euclidean space.
Both $\langle H_{M}\rangle$ and $\langle H_{a}\rangle$ are independent of scale and renormalization schemes. In contrast, the quark energy $\langle H_{E}^{(\mu)}\rangle$ and gluon field energy $\langle H_{g}^{(\mu)}\rangle$ are dependent on both scale and renormalization. 

Thus, the renormalized quark and gluon energy are derived as 
\begin{align}
\left\langle H_{E}^{R}\right\rangle  & =\frac{3}{4}\langle x\rangle_{q}^{R}M-\frac{3}{4}\left\langle H_{M}\right\rangle ,\nonumber \\
\langle H_{g}^{R}\rangle & =\frac{3}{4}\langle x\rangle_{g}^{R}M,
\end{align}
where $\langle x\rangle_{q}^{R}$ and $\langle x\rangle_{g}^{R}$ are the renormalized momentum fractions of quarks and gluons, respectively. These fractions satisfy the relation $\langle x\rangle_{g}^{R}=1-\langle x\rangle_{q}^{R}$~\citep{QCDSF:2012mkm}. Following Ref.~\citep{Sun:2020pda}, we
could also define the total valence charm quark contribution as
\begin{equation}
\langle H_{q}^{R}\rangle=\langle H_{E}^{R}\rangle+\langle H_{M}\rangle=\frac{3}{4}\langle x\rangle_{q}^{R}M+\frac{1}{4}\langle H_{M}\rangle.\label{the total Hamiltonian of the quark field}
\end{equation}
In combination with the trace sum rule~\citep{Shifman:1978zn} 
\begin{eqnarray}
M=\langle T_{\mu}^{\mu}\rangle=\langle H_{M}\rangle+\langle H_{a}\rangle,\label{the trace sum rule}
\end{eqnarray}
separate part of the mass decomposition will be gained, provided with the calculation of the hadron mass M, the quark condensate contribution
$\langle H_{M}\rangle$ and the quark energy contribution $\langle H_{E}\rangle$.

\subsection{Two-point and three-point functions}\label{Two point and three point function}

The components of the mass decomposition can be extracted from the
corresponding two-point and three-point correlation functions. To construct
the correlation function for the triply charmed baryon, similar to
the Omega baryon~\citep{CLQCD:2015bgi}, we use the operator for the
triply charmed baryon as:
\begin{eqnarray}
O^{\mu}_{\textcolor{red}{\gamma}} (\vec{x},t)=\epsilon^{abc}[\psi_{\alpha}^{a}(\vec{x},t)^{T}(C\gamma^{\mu})_{\alpha\beta}\psi_{\beta}^{b}(\vec{x},t)]\psi_{\gamma}^{c}(\vec{x},t),
\end{eqnarray}
where $C=\gamma_{2}\gamma_{4}$ is the $C$-parity operator; $\alpha$,
$\beta$, $\gamma$ represent the Dirac indices; $a$, $b$, $c$
are the color indices; and $T$ is the transpose operator. To project
onto a definite parity, we use the following parity projection operator:
\begin{eqnarray}
P_{\pm}=\frac{1}{2}(1\pm\gamma_{4}).
\end{eqnarray}
Additionally, to project onto the triply charmed baryon with a definite
spin, we use the following spin projection operators~\citep{EuropeanTwistedMass:2008pab}:
\begin{equation}
\begin{cases}
P_{\frac{3}{2}}^{\mu\nu}=\delta^{\mu\nu}-\frac{1}{3}\gamma^{\mu}\gamma^{\nu},\\
P_{\frac{1}{2}}^{\mu\nu}=\frac{1}{3}\gamma^{\mu}\gamma^{\nu}.
\end{cases}
\end{equation}
In our study, only the spatial components of the triply charmed baryon
operator are considered. Therefore, the baryon operator with a definite
$J^{P}$ quantum number can be expressed as:

\begin{equation}
\begin{aligned} & O^{i}(\vec{x},t)=(P_{\pm})_{\rho\rho^{\prime}}\\
 & \times\sum_{j}(P_{J})_{\rho^{\prime}\gamma}^{ij}\epsilon^{abc}\left[\psi_{\alpha}^{a}(\vec{x},t)^{T}(C\gamma^{j})_{\alpha\beta}\psi_{\beta}^{b}(\vec{x},t)\right]\psi_{\gamma}^{c}(\vec{x},t).
\end{aligned}
\end{equation}

The hadron mass $M$ can be obtained from the two-point correlation
function 
\begin{eqnarray}
C_{2}(t) & = & \sum_{\vec{x}}\langle O(\vec{x},t)O^{\dagger}(\vec{0},0)\rangle\nonumber \\
 & = & \sum_{n}Z_{n}^{2}e^{-M_{n}t}\xrightarrow{t\to\infty}Z_{0}^{2}e^{-M_{0}t},
\end{eqnarray}
Here, $M_{0}$ represents the ground state hadron mass, and $Z_{0}$
is the overlap matrix element between the ground state hadron and
the hadron operator. Hadronic matrix elements, such as the quark content
$\langle H_{M}\rangle$ and the quark momentum fraction $\langle x\rangle_{q}$
, can be extracted from the three-point function 
\begin{eqnarray}
C_{3}(t,t^{\prime},J,\hat{O}) & = & \sum_{\vec{x},\vec{y},t^\prime}\langle O(\vec{x},t)J(\vec{y},t^{\prime})O^{\dagger}(\vec{0},0)\rangle,\label{three point function} \nonumber \\
\xrightarrow{t \gg t^\prime \gg 0,t\to\infty} &  & Z_{0}^{2}e^{-M_{0}t}\langle\Omega_{ccc}|J(\vec{0})|\Omega_{ccc}\rangle
\end{eqnarray}
where $J(\vec{y},t^{\prime})$ refers to the current operator. Here, we only considered the contribution of the
valence charm quark. For the quark content $\langle H_{m}\rangle$,
the corresponding current operator is 
\begin{eqnarray}
\hat{H}_{M}(\vec{y},t^{\prime})=m_{c}\bar{\psi}^{(c)}(\vec{y},t^{\prime})\psi^{(c)}(\vec{y},t^{\prime}),
\end{eqnarray}
where $m_{c}$ refers to the bare charm quark mass. For the charm
quark momentum fraction $\langle x\rangle_{q}$, the current operator
is 
\begin{equation}
\hat{x}_{q}(\vec{y},t^{\prime})=\frac{1}{2}\bar{\psi}^{(c)}(\vec{y},t^{\prime})(\gamma_{4}\overleftrightarrow{D}_{4}-\frac{1}{3}\gamma_{i}\overleftrightarrow{D}_{i})\psi^{(c)}(\vec{y},t^{\prime}).
\end{equation}

\section{Numerical details} \label{Numerical details}
In this calculation, we used the 2+1 flavor domain wall fermion and
Iwasaki gauge action configurations provided by the RBC/UKQCD collaboration \citep{RBC:2010qam, Mawhinney:2019cuc}. Table~\ref{configuration parameter} presents the parameters of these gauge ensembles. For the valence charm quark, we employ the overlap fermion with exact chiral symmetry on the lattice, which ensures that the valence charm quark mass matrix $\langle H_{M}\rangle$ is renormalization scale and scheme independent~\citep{Liu:2013yxz}.
The valence charm quark mass adopted in both ensembles follows the same tuning procedure as in Ref. \citep{Sun:2020pda}, where the physical $J/\psi$ mass was used as the matching criterion.

\begin{table}
\caption{The parameters for the configurations~\citep{Sun:2020pda}}
\label{configuration parameter} \centering %
\begin{ruledtabular}
\begin{tabular}{cccccc}
\text { ensemble }  & $L^{3}\times T$  & $a(\mathrm{fm})$  & $m_{\pi}(\mathrm{MeV})$  & $m_{c}a$  & $N_{\mathrm{cfg}}$ \tabularnewline
\hline 
\text { 32I }  & $32^{3}\times64$  & $0.0828(3)$  & 300  & 0.493  & 305 \tabularnewline
\text { 48If }  & $48^{3}\times96$  & 0.0711(3)  & 278  & 0.410  & 205 \tabularnewline
\end{tabular}
\end{ruledtabular}
\end{table}

To extract the hadron mass, we directly fit the two-point correlation
function. Considering the unphysical oscillatory behavior introduced
by the Domain Wall fermion~\citep{Liang:2013eoa}, we use the following
fitting function for the two-point function:
\begin{equation}
C_{2}(t)=A_{0}e^{-Mt}\left(1+A_{1}e^{-\delta mt}\right)+W(-1)^{t}e^{-\widetilde{M}t}.    
\end{equation}
In this expression, $e^{-\delta mt}$ is used to absorb the contributions
from the excited states, and $e^{-\widetilde{M}t}$ is the oscillating
term. The parameters $M$, $\delta m$, $\widetilde{M}$, $W$, $A_{0}$ and $A_{1}$
are determined through the fitting process. This approach allows us
to account for and mitigate the effects of oscillations in the data
when determining the hadron mass.

\begin{figure*}[htbp]
\centering 
\subfigure{ %
\begin{minipage}[c]{7cm} %
 \centering \includegraphics[width=7cm]{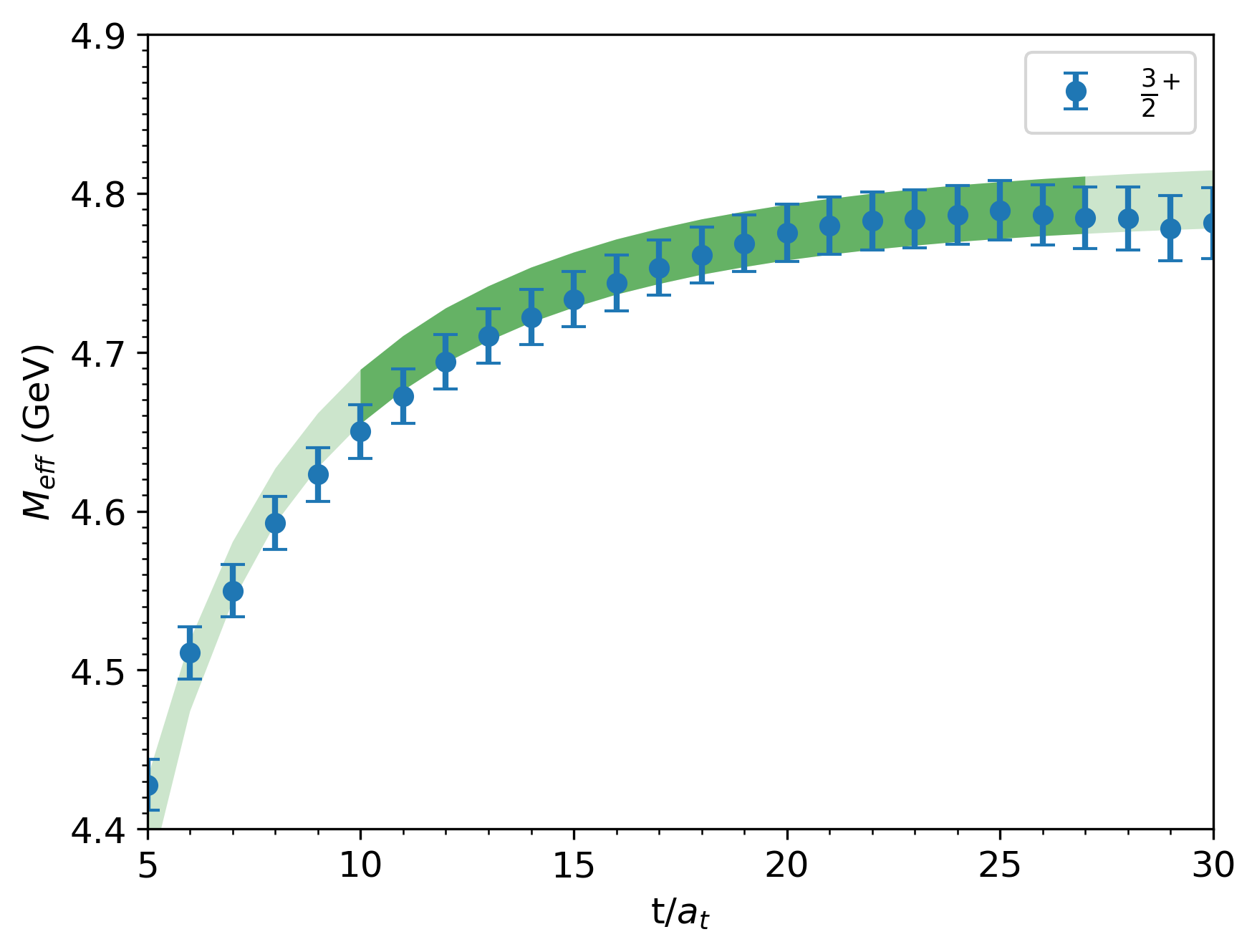}
%
\end{minipage}}
\subfigure{ %
\begin{minipage}[c]{7cm} %
 \centering \includegraphics[width=7cm]{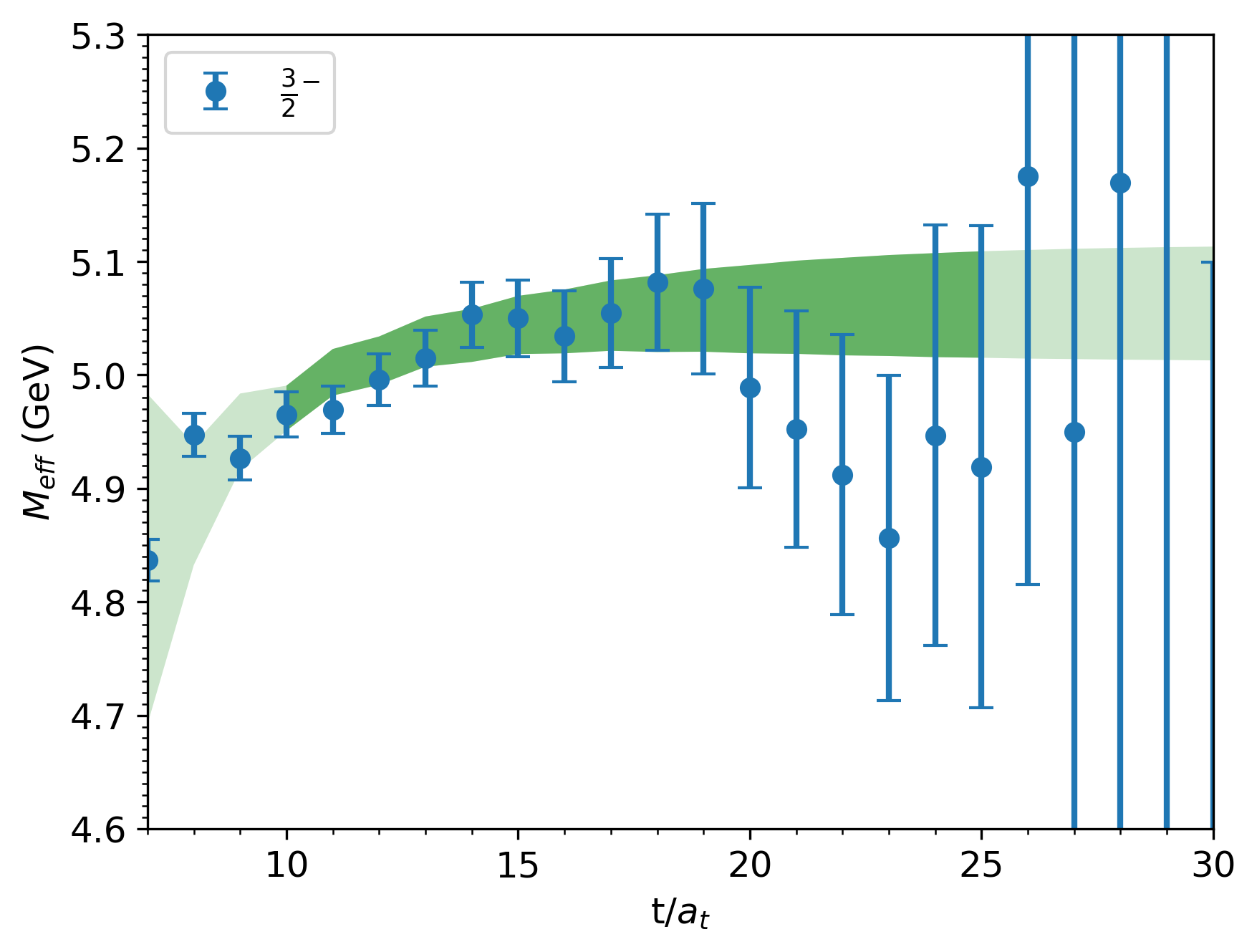}
\end{minipage}} 
\subfigure{ %
\begin{minipage}[c]{7cm} %
 \centering \includegraphics[width=7cm]{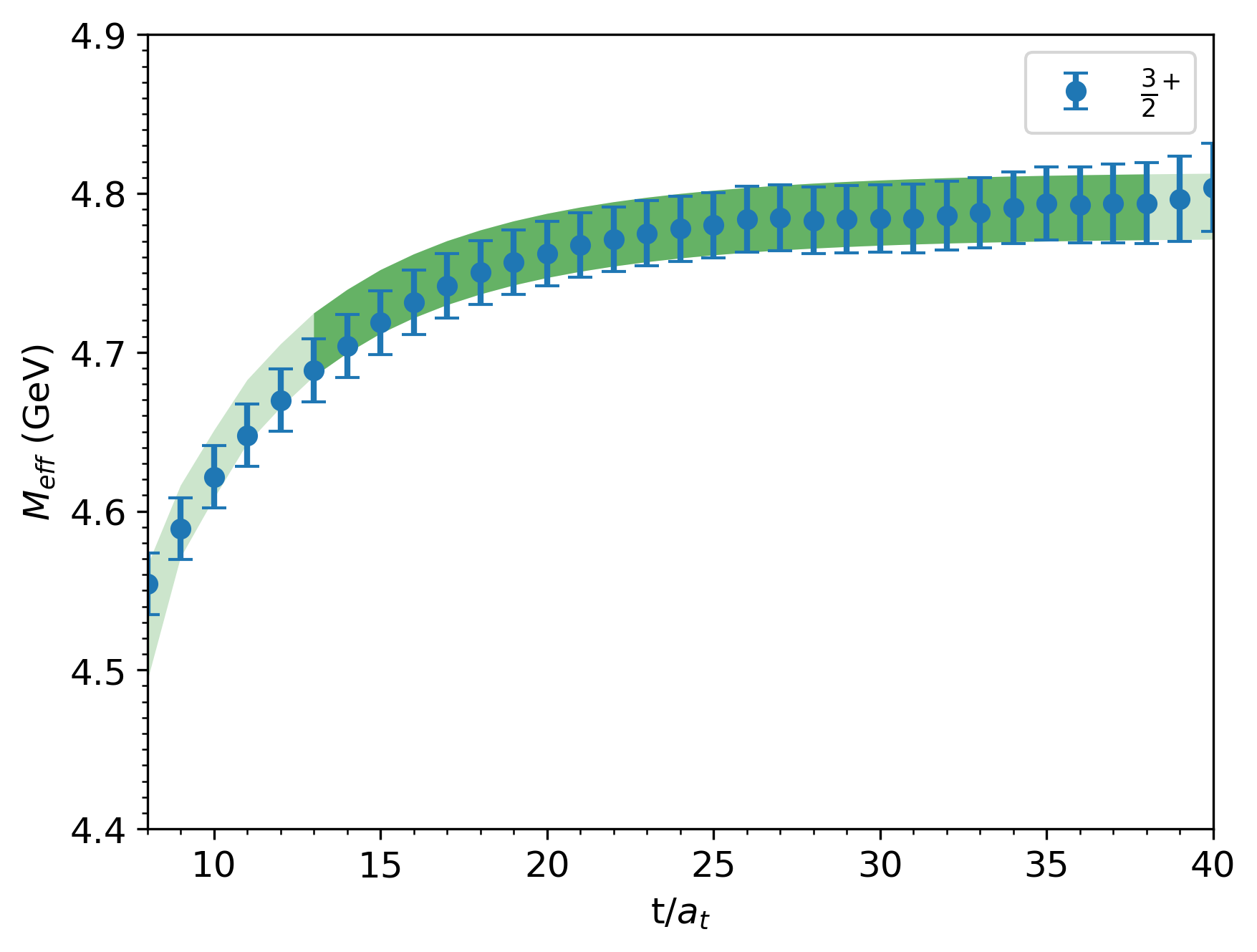}
\end{minipage}}
\subfigure{ %
\begin{minipage}[c]{7cm} %
 \centering \includegraphics[width=7cm]{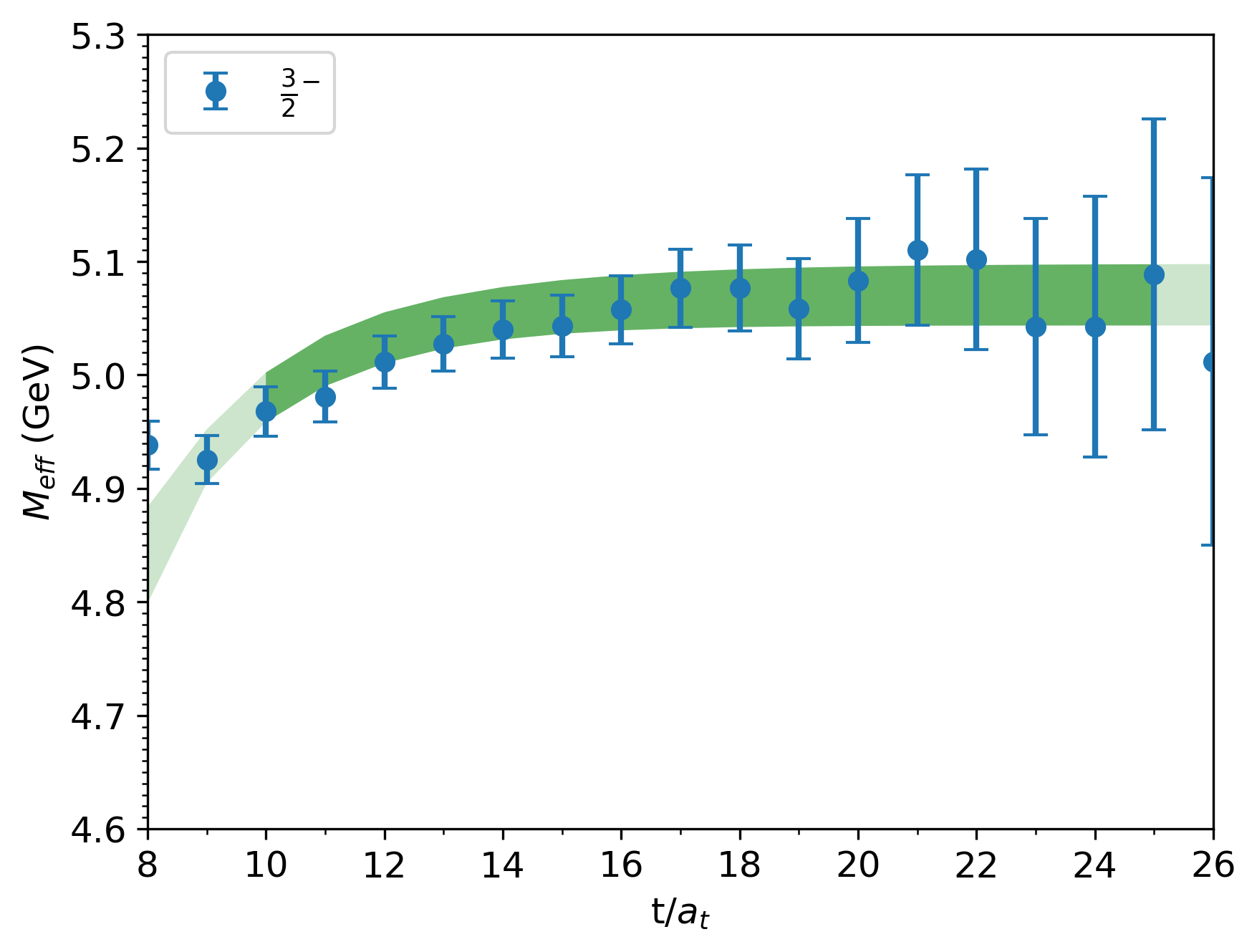}
\end{minipage}} 
\caption{Effective mass $m_{{\rm eff}}={\rm ln}\left(\frac{C_{2}(t+1)}{C_{2}(t)}\right)$
for two triply charmed baryon states on $32^{3}\times64$ (top) and
$48^{3}\times96$ (bottom) configurations.}
\label{Effective mass} 
\end{figure*}

The effective masses of the two triply charmed baryons with quantum
numbers $J^{P}=\frac{3}{2}^{+}$ and $J^{P}=\frac{3}{2}^{-}$, obtained
from two different lattice configurations, are depicted in Fig.\ref{Effective mass}. In the figure, the dark color bands represent our fitting
range, while the light color bands indicate the extrapolation results.
The fitted masses and $\delta m$ of the two different triply charmed baryons are shown
in Table.~\ref{hadron mass}. Due to the
fact that we have only two lattice spacings and the results on the two ensembles are consistent within errors, we perform a constant extrapolation to obtain the results at the continuum limit.

\begin{table}
\caption{The hadron mass for two triply charmed baryon states on $32^{3}\times64$
(32I) and $48^{3}\times96$ (48If) configurations, along with the
fitting range $[t_{min}-t_{max}]$ and $\chi^{2}/d.o.f$. The results of constant extrapolation are also shown in the last row. }
\label{hadron mass} \centering %
\begin{ruledtabular}

\begin{tabular}{cccccc}
ensemble  & $J^{P}$  & M(GeV)  & $\delta m$(GeV) & $[t_{min}-t_{max}]$  & $\chi^{2}/d.o.f$ \tabularnewline
\hline 
32I  & $\frac{3}{2}^{+}$  & 4.804(20) & 0.300(52)  & 10-27  & 0.76 \tabularnewline
~~~~~ & $\frac{3}{2}^{-}$  & 5.064(51) &0.64(64) & 10-25  & 1.4 \tabularnewline
\hline 
48If  & $\frac{3}{2}^{+}$  & 4.793(21)  &0.439(58) & 13-38  & 0.42 \tabularnewline
~~~~~ & $\frac{3}{2}^{-}$  & 5.071(27)  &1.14(36) &10-25  & 0.85 \tabularnewline
\hline 
$\infty$  & $\frac{3}{2}^{+}$  & 4.799(14)  &  &   & 
\tabularnewline
~~~~~ & $\frac{3}{2}^{-}$  & 5.069(24)  &   &  & 
\end{tabular}

\end{ruledtabular}
\end{table}

To obtain the charmness content $\langle H_{M}\rangle$ and the charm
quark momentum fraction $\langle x\rangle_{q}$, we need to calculate
the corresponding three-point correlation functions. We employ the Feynman-Hellmann inspired method to compute the three-point function as done in Ref.~\citep{Chang:2018uxx,Sun:2020pda} and readers are referred to Ref.~\citep{Bouchard:2016heu} for more details. The corresponding
current-summed three-point function is given as
\begin{align}
C^{(3)}(t & ,J,\mathcal{O})=\sum_{t^{\prime}=0}^{T-1}\langle0|T\left\{ \mathcal{O}(t)\mathcal{J}\left(t^{\prime}\right)\mathcal{O}^{\dagger}(0)\right\} |0\rangle \nonumber \\
 & =\sum_{y,c}\left\langle \Gamma G_{c}^{J}(\vec{y},t;0)G(\vec{y},t;0)G(\vec{y},t;0)\right\rangle , \nonumber \\
 & =\sum_{t^{'}=1}^{t_{}-1}\sum_{n,m}\langle0|\mathcal{O}(t)|n\rangle\langle n|\mathcal{J}\left(t^{\prime}\right)|m\rangle\langle m|\mathcal{O}^{\dagger}(0)|0\rangle \nonumber \\
 & +II+III+VI \nonumber \\
 & =\sum_{n}\left[(t-1)Z_{n}J_{nn}Z_{n}^{\dagger}+d_{n}\right]e^{-E_{n}t} \nonumber \\
 & +\sum_{n\neq m}Z_{n}J_{nm}Z_{m}^{\dagger}\frac{e^{-E_{n}t}e^{\frac{\Delta_{nm}}{2}}-e^{-E_{m}t}e^{\frac{\Delta_{mn}}{2}}}{e^{\frac{\Delta_{mn}}{2}}-e^{\frac{\Delta_{nm}}{2}}}.
\end{align}
Here, the sum over $c$ represents the various possible contractions of the current coupling to a quark propagator and $G_{c}^{J}$ denotes the Feynman-Hellmann propagator
\begin{equation}
  G_{c}^{J}(\vec{y},t;0)=\sum_{\vec{x},t^{\prime}}G(\vec{y},t;\vec{x},t^{\prime})J(\vec{x},t^{\prime})G(\vec{x},t^{\prime};0).  
\end{equation}
 $\Gamma$ represents the product of the initial and final state
$\gamma$ matrices. For simplicity, the contraction over color indices
is omitted.  II, III, and IV denote the contributions from regions of $t<t'<T$,
$t'=0$, and $t'=t$, respectively. And they are absorbed in the $d_{n}$ term despite a $e^{-E_{n}t}$ factor. 
$Z_{n}$ represents the overlap factor. $J_{nn}$ denotes the hadronic
matrix element. $\Delta_{nm}$ is defined as $\Delta_{nm}\equiv E_{m}-E_{n}$.

The ratio of the three-point function to the corresponding
two-point function is defined as
\begin{align}
R(t) & =\frac{C^{(3)}(t)}{C^{(2)}(t)} \nonumber \\
 & \simeq(t-1)J_{00}+\frac{d_{n}}{|Z_{0}|^{2}}+J_{10}\frac{Z_{1}}{Z_{0}}\frac{e^{\frac{\Delta_{10}}{2}}}{e^{\frac{\Delta_{10}}{2}}-e^{\frac{\Delta_{01}}{2}}} \nonumber \\
 & +\sum_{n=1}C_{n}(t-1)e^{-\Delta_{n0}t}+D_{n}e^{-\Delta_{n0}t}.
\end{align}
In the second step, we assume that $C^{(2)}(t)$ is primarily contributed by the ground state,
and incorporate some of the coefficients related to the excited states into $C_n$ and $D_n$, for the sake of simplicity.
We can derive the matrix element $J_{00}$ from the derivative of R(t)
\begin{equation}
    \partial_{t}R(t) \equiv R(t-1)-R(t)\xrightarrow{t\to\infty}J_{00}.
\end{equation}
Although the subtraction procedure may introduce larger relative
statistical uncertainties, this method efficiently generates results
for multiple all sink times. It avoids the computational cost of repeatedly
calculating propagators for each sink time (as in traditional sink-sequential
approaches). The increased data density across time separations enables
better control of excited-state contamination and enhances the accuracy
of extracting the desired hadronic matrix elements.
The lattice correlator with all sink times available enables
better control of excited-state contamination (at sink) and thus better extraction of the desired matrix elements. The final numerical results are all of decent precision.

The effective hadronic matrix elements can be obtained from the difference of R(t) as
\begin{align}
    \langle H_{M}\rangle(t) &=R(t,\hat{H}_{M},\hat{Q})-R(t-1,\hat{H}_{M},\hat{Q}) \nonumber \\
M\langle x\rangle_{q}(t) &=R(t,\hat{x},\hat{Q})-R(t-1,\hat{x},\hat{Q}).
\end{align}
And then the hadronic matrix elements $\langle H_{M}\rangle$ and $M\langle x_{q}\rangle$ can be fitted using the formulas as 
\begin{align}
\langle H_{M}\rangle(t)&=\langle H_{M}\rangle+A_{1}^{\prime}e^{-\delta mt}+tA_{2}^{\prime}e^{-\delta mt}  \nonumber \\
M\langle x_{q}\rangle(t)&=M\langle x_{q}\rangle+A_{3}^{\prime}e^{-\delta mt}+tA_{4}^{\prime}e^{-\delta mt},
\end{align}
where $\delta m$, $H_M$, $x_q$, $A_{1}^{\prime},A_{2}^{\prime},A_{3}^{\prime},$ and $A_{4}^{\prime}$
are fitting parameters, and the exponential terms are used to absorb the contribution from excited states.

\begin{figure*}[htbp]
\centering \subfigure{ %
\begin{minipage}[c]{7cm}%
 \centering \includegraphics[width=7cm]{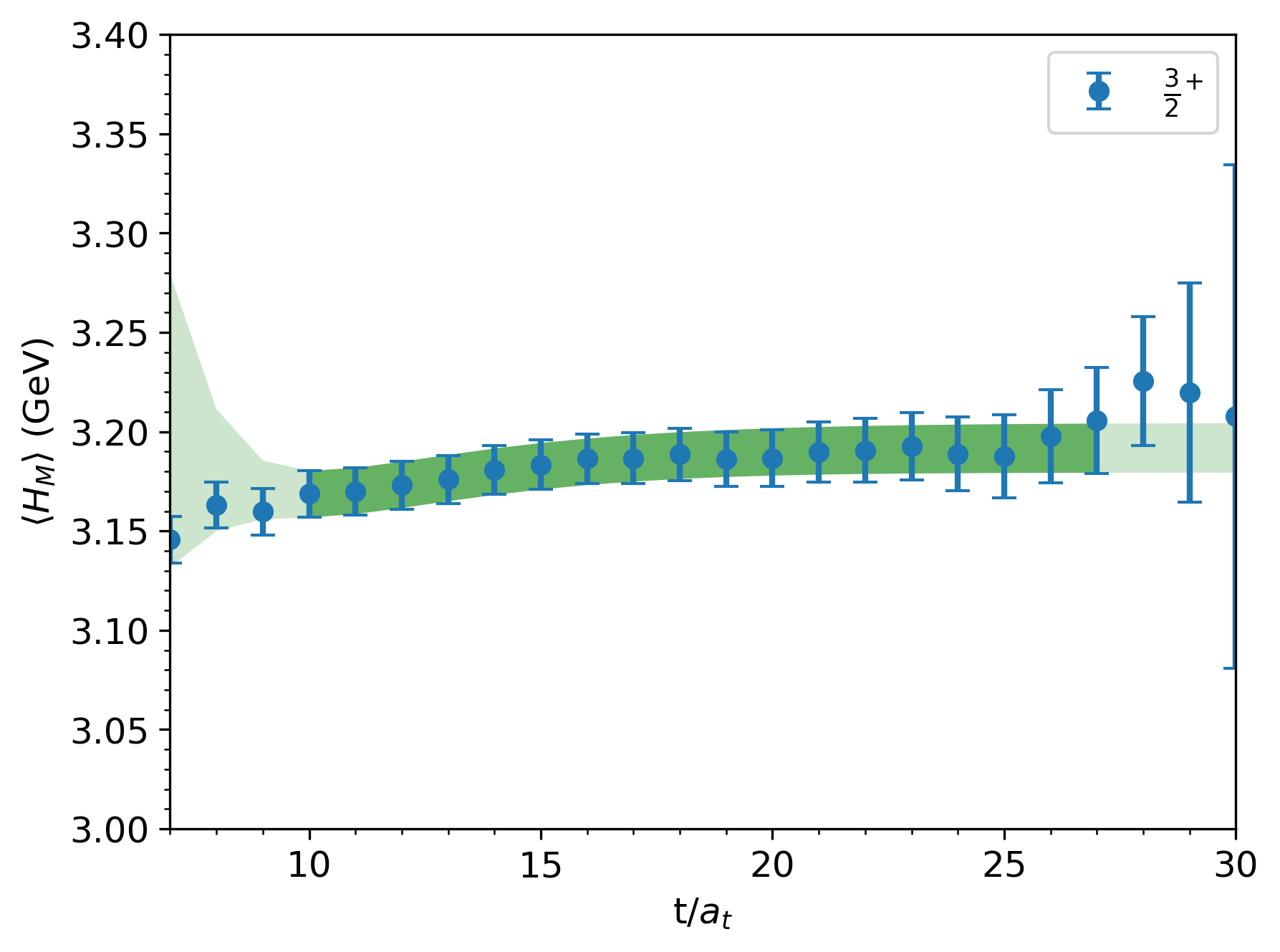}
\end{minipage}}\subfigure{ %
\begin{minipage}[c]{7cm}%
 \centering \includegraphics[width=7cm]{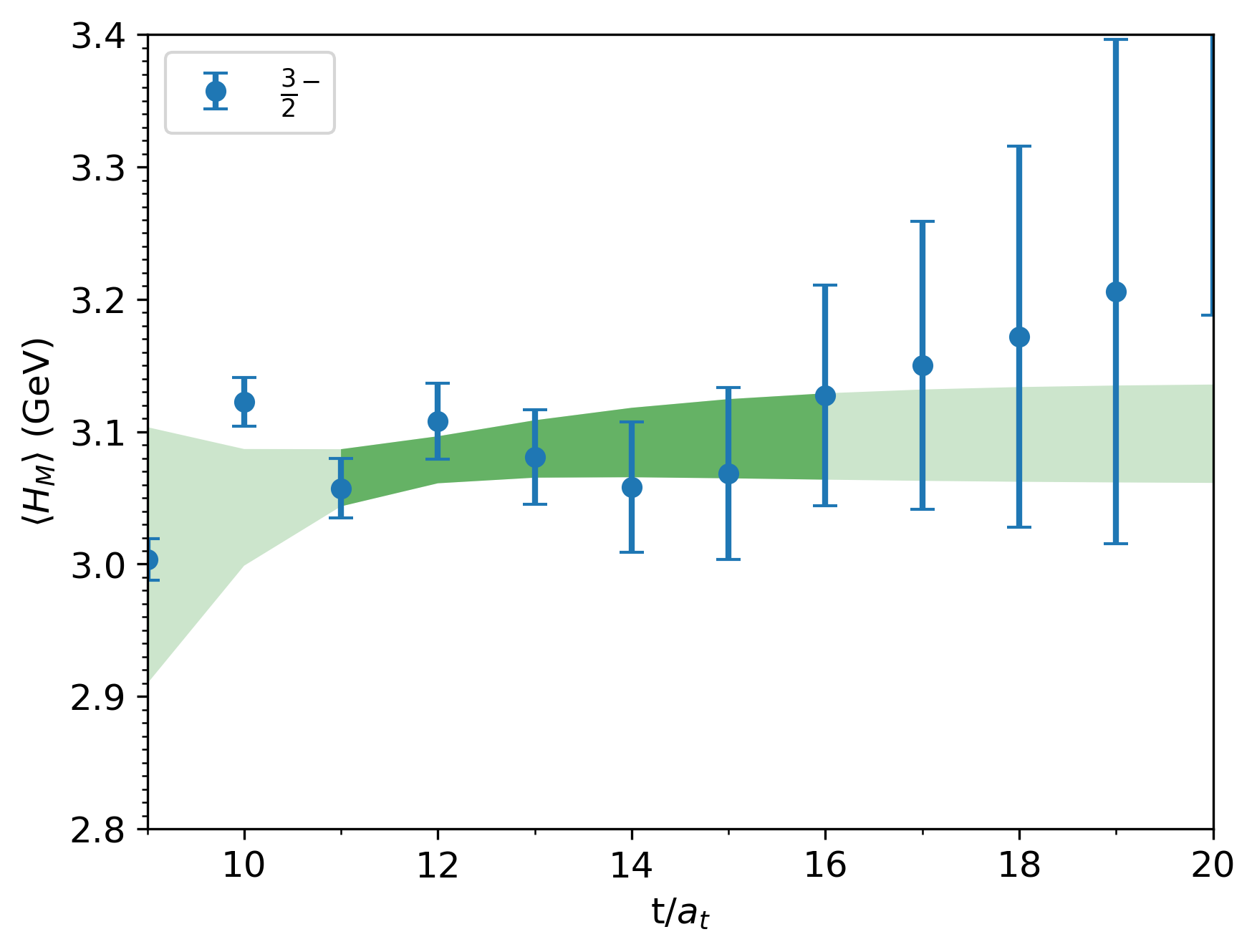}
\end{minipage}} \subfigure{ %
\begin{minipage}[c]{7cm}%
 \centering \includegraphics[width=7cm]{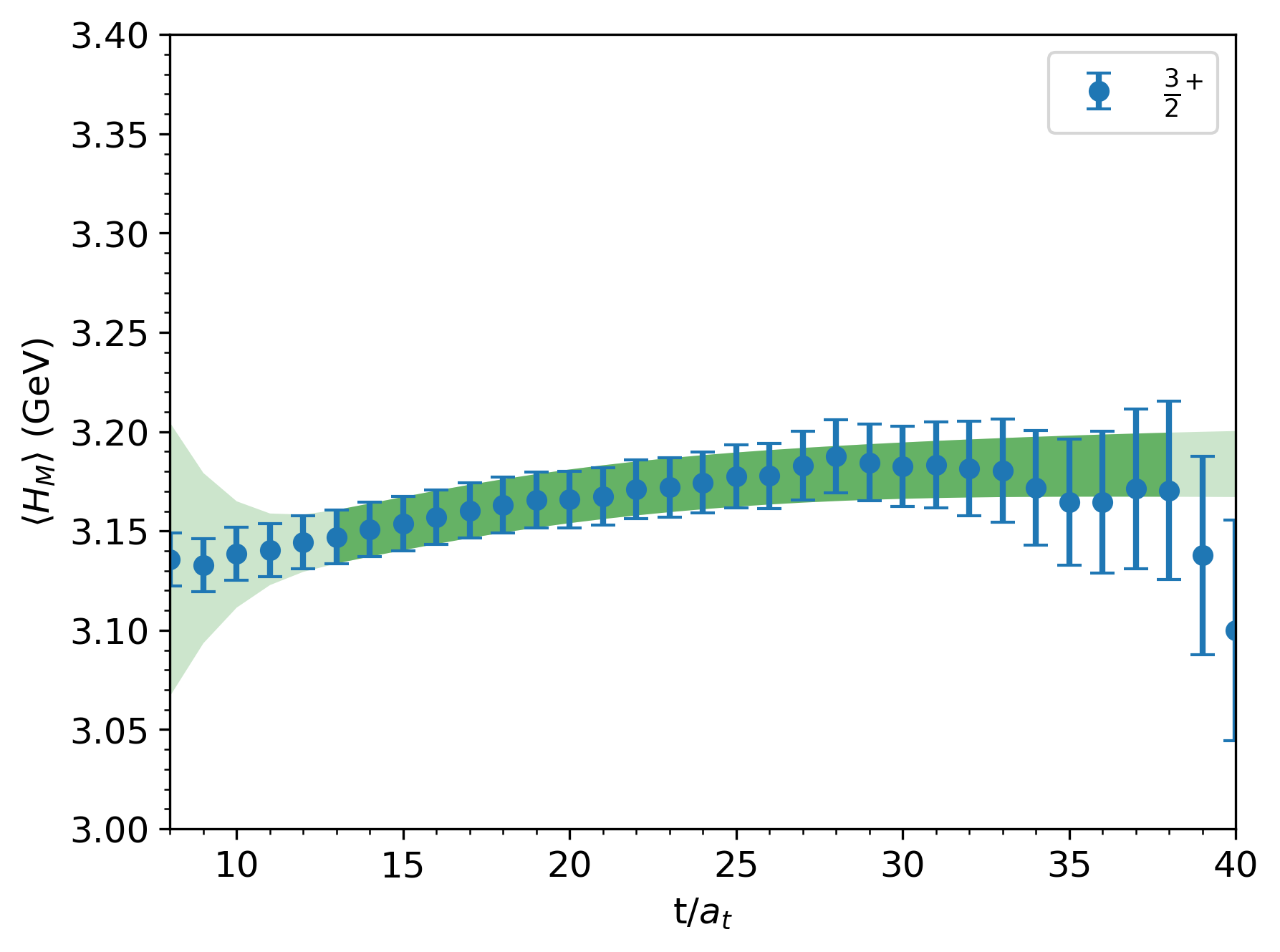}
\end{minipage}}\subfigure{ %
\begin{minipage}[c]{7cm}%
 \centering \includegraphics[width=7cm]{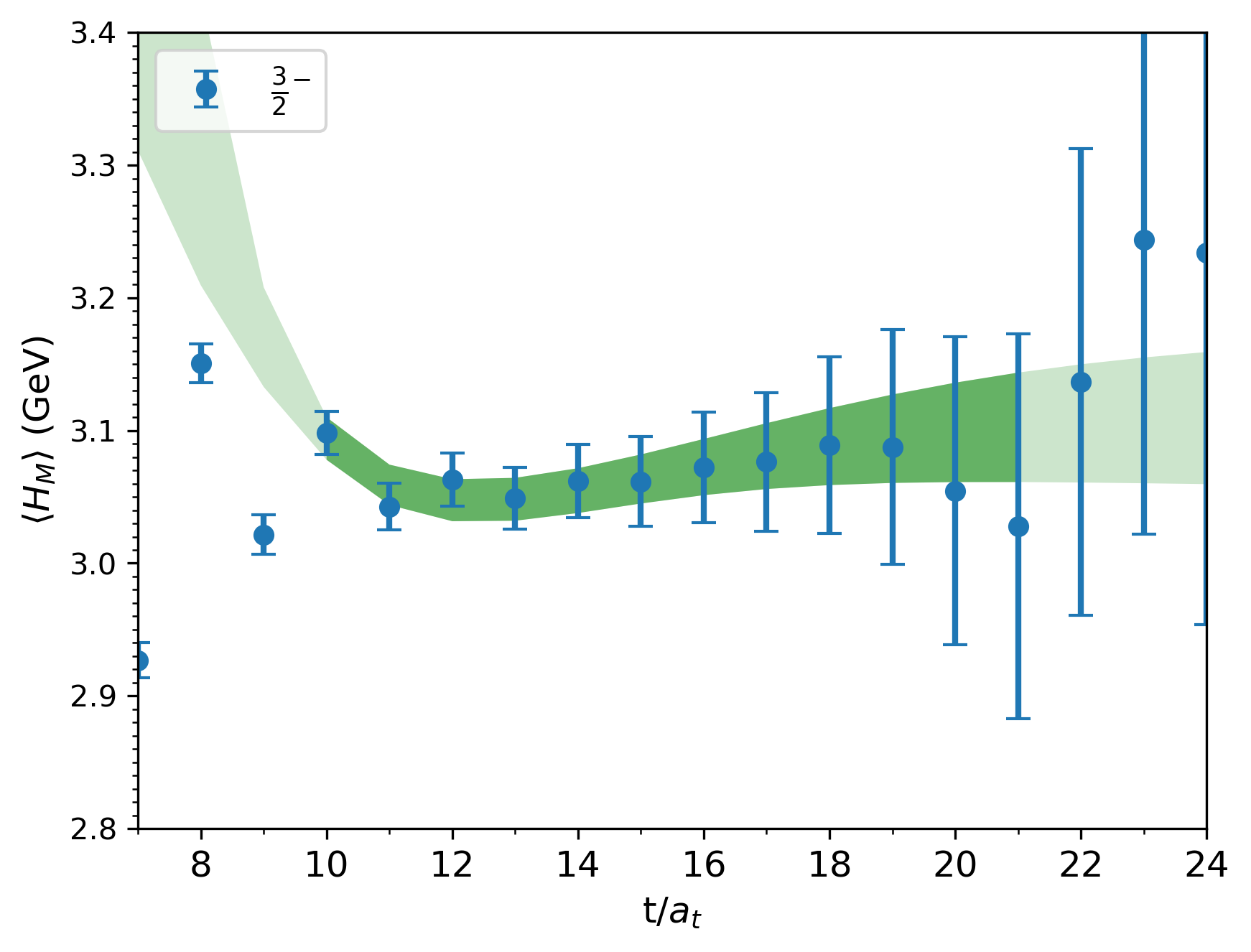}
\end{minipage}} \caption{Effective matrix elements of  the charmness content $\langle H_{M}\rangle$
for two triply charmed baryon states on $32^{3}\times64$ (top) and
$48^{3}\times96$ (bottom) configurations.}
\label{Effective matrix element of charmed quark mass} 
\end{figure*}

\begin{figure*}[htbp]
\centering \subfigure{ %
\begin{minipage}[c]{7cm}%
 \centering \includegraphics[width=7cm]{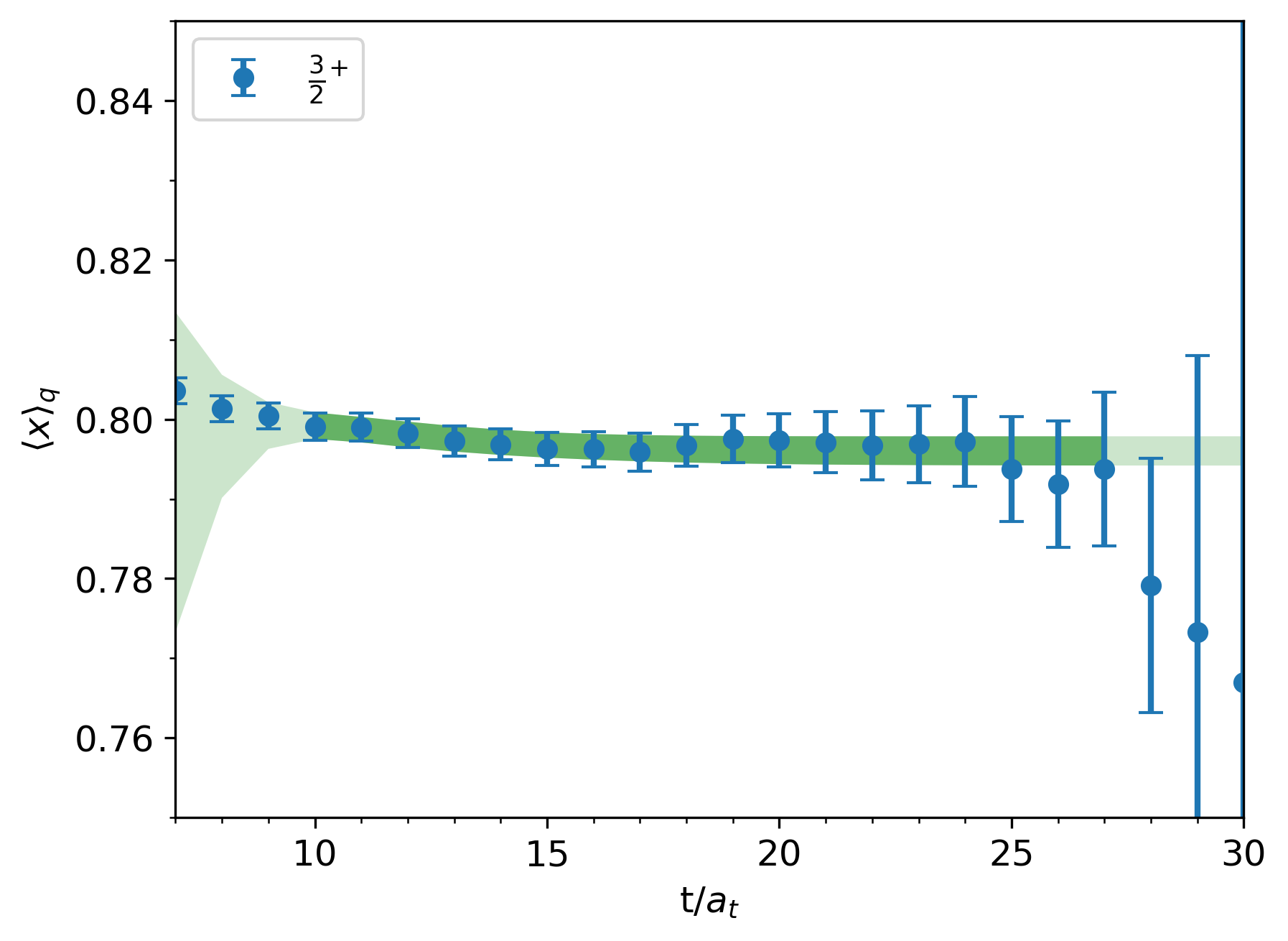}
\end{minipage}}\subfigure{ %
\begin{minipage}[c]{7cm}%
 \centering \includegraphics[width=7cm]{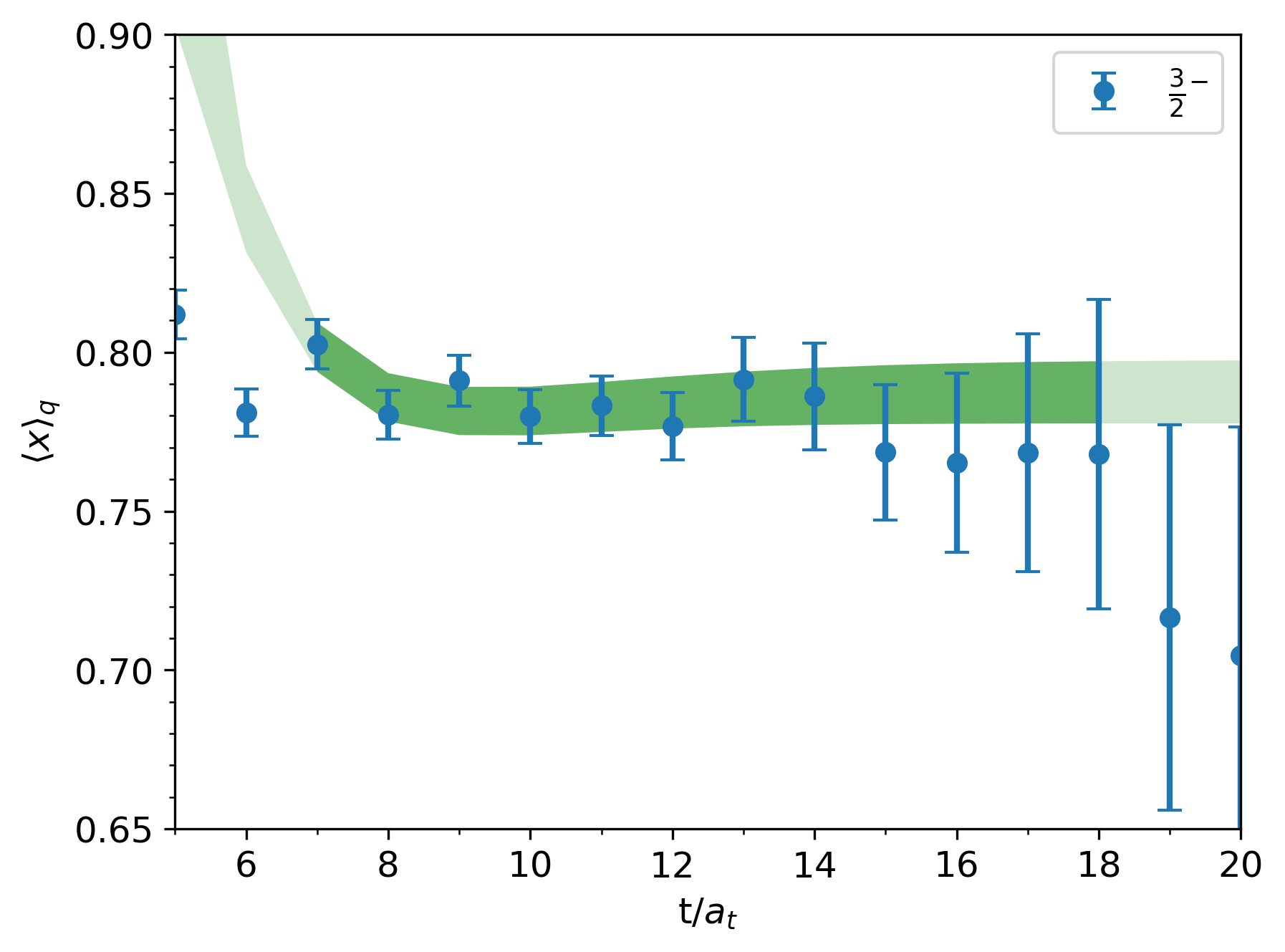}
\end{minipage}} \subfigure{ %
\begin{minipage}[c]{7cm}%
 \centering \includegraphics[width=7cm]{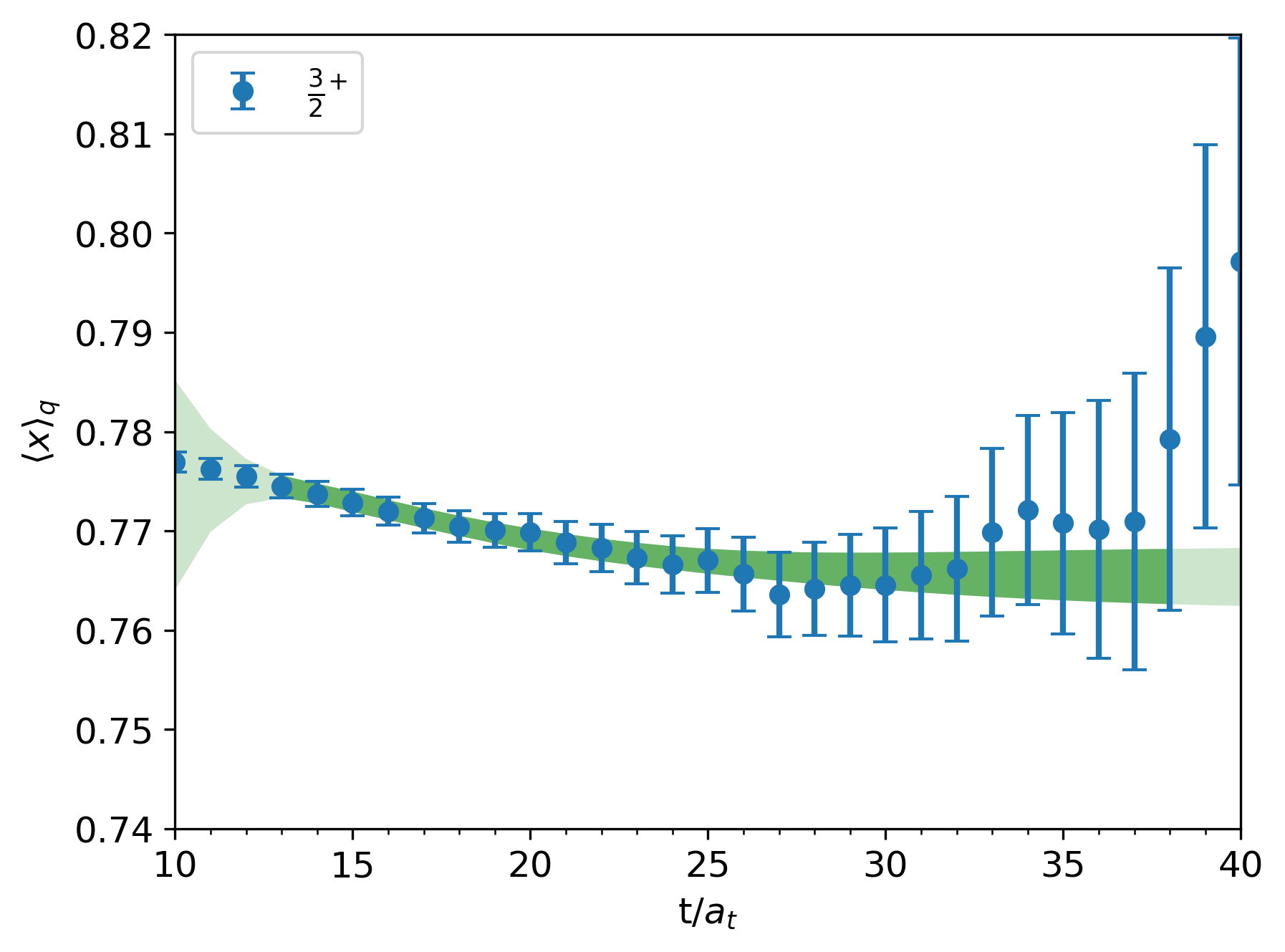}
\end{minipage}}\subfigure{ %
\begin{minipage}[c]{7cm}%
 \centering \includegraphics[width=7cm]{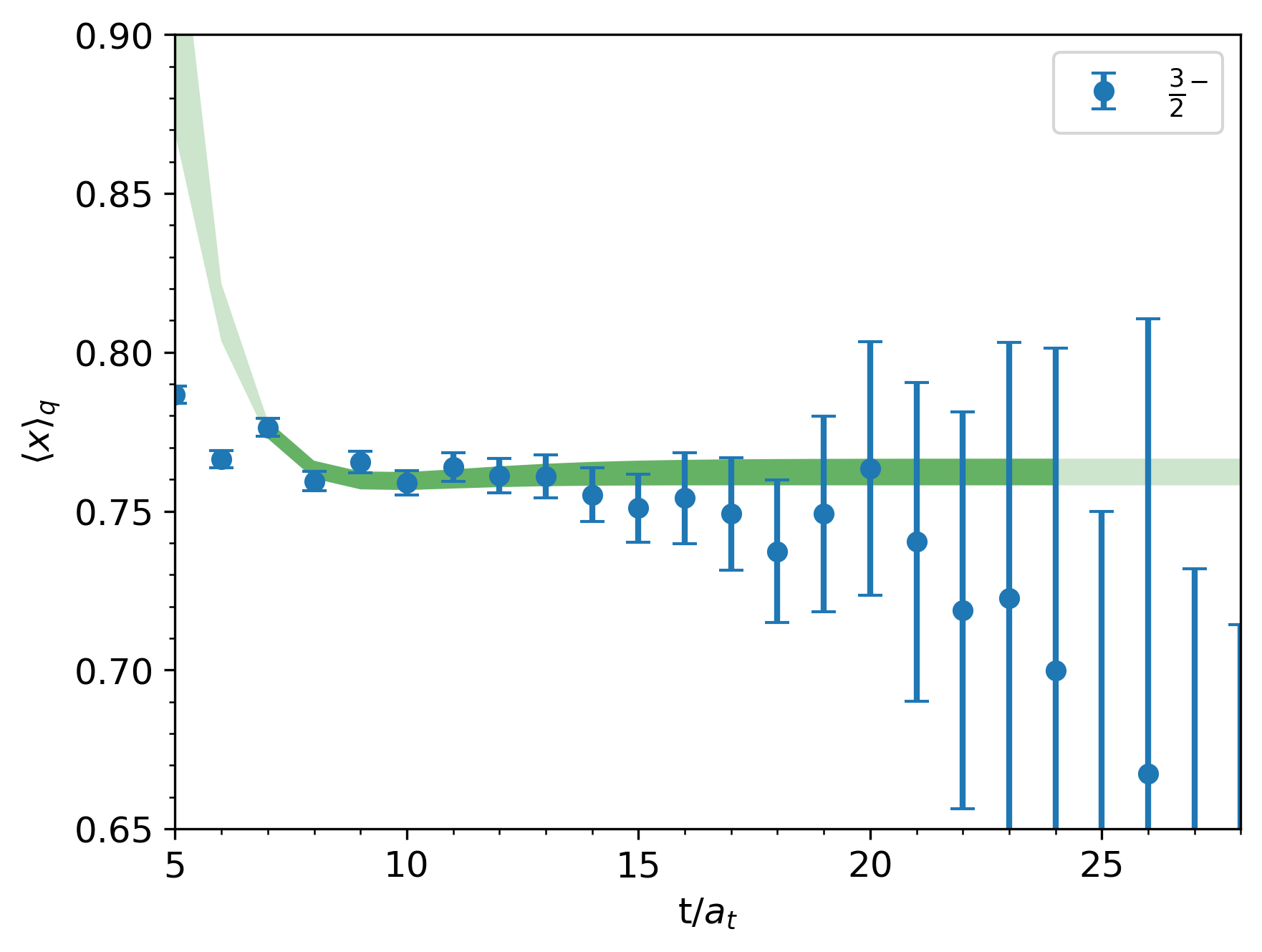}
\end{minipage}} \caption{Effective matrix elements of valence charmed quark momentum fraction
$\langle x\rangle_{q}$ for two triply charmed baryon states on $32^{3}\times64$
(top) and $48^{3}\times96$ (bottom) configurations.}
\label{Effective matrix element of charmed quark momentum fraction} 
\end{figure*}

The fitting of effective matrix elements of the charmness content $\langle H_{M}\rangle$ and the valence charmed quark momentum
fraction $\langle x\rangle_{q}$ are shown in Fig.~\ref{Effective matrix element of charmed quark mass}
and Fig.~\ref{Effective matrix element of charmed quark momentum fraction}, respectively.
The dark color bands represent our fitting range, while the light color
bands show the extrapolation results. The fitted
results of the corresponding hadron matrix elements are shown in Table~\ref{charmed quark mass}
and Table~\ref{charmed quark momentum fraction}. Our results reveal
that, on two different configurations, the charmness content of the
orbital excited state $\frac{3}{2}^{-}$ is slightly smaller than
that of the ground state $\frac{3}{2}^{+}$.

\begin{table}
\caption{The charmness content $\langle H_{M}\rangle$ for two triply charmed
baryon states on $32^{3}\times64$ (top) and $48^{3}\times96$ (bottom)
configurations, along with the corresponding fitting range $[t_{min}-t_{max}]$
and $\chi^{2}/d.o.f$ .}
\label{charmed quark mass} \centering %
\begin{ruledtabular}
\begin{tabular}{cccccc}
ensemble  & $J^{P}$  & $\langle H_{M}\rangle$(GeV)  &  $[t_{min}-t_{max}]$  & $\chi^{2}/d.o.f$ \tabularnewline
\hline 
32I  & $\frac{3}{2}^{+}$  & 3.192(13) &  10-27 & 0.062\tabularnewline
~~~~~ & $\frac{3}{2}^{-}$ & 3.098(38)&  11-16  & 0.38 \tabularnewline
\hline 
48If  & $\frac{3}{2}^{+}$  & 3.185(19)&  13-38  & 0.089 \tabularnewline
~~~~~ & $\frac{3}{2}^{-}$  & 3.114(60)&  10-21  & 0.33 \tabularnewline
\end{tabular}
\end{ruledtabular}
\end{table}

\begin{table}
\caption{The charmed quark momentum fraction $\langle x\rangle_{q}$ for two
triply states on $32^{3}\times64$ (top) and $48^{3}\times96$ (bottom)
configurations, along with the corresponding fitting range $[t_{min}-t_{max}]$
and $\chi^{2}/d.o.f$ .}
\label{charmed quark momentum fraction} \centering %
\begin{ruledtabular}
\begin{tabular}{cccccc}
ensemble  & $J^{P}$  & $\langle x\rangle_{q}$ &  $[t_{min}-t_{max}]$  & $\chi^{2}/d.o.f$ \tabularnewline
\hline 
32I  & $\frac{3}{2}^{+}$  & 0.7960(18) &  10-27  & 0.096 \tabularnewline
 & $\frac{3}{2}^{-}$  & 0.7878(99)     &  7-18  & 1.7\tabularnewline
\hline 
48If  & $\frac{3}{2}^{+}$  & 0.7653(33)&  13-38  & 0.12 \tabularnewline
 & $\frac{3}{2}^{-}$  & 0.7623(42)     &  7-24  & 1.1 \tabularnewline
\end{tabular}
\end{ruledtabular}
\end{table}

\section{Discussion} \label{Discussion}

\subsection{Mass spectrum}\label{Mass spectrum}

\begin{figure}
    \centering
 \includegraphics[width=7cm]{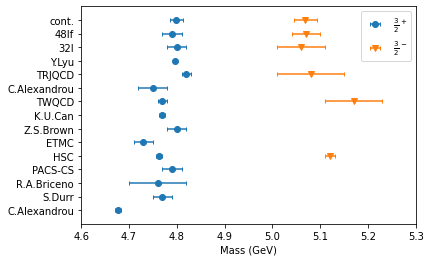}
    \caption{Masses of $\frac{3}{2}^{+}$
and $\frac{3}{2}^{-}$  triply charmed baryons on lattice QCD.}
    \label{fig:Mass}
\end{figure}

Before delving into the mass decomposition of the triply charmed baryon,
let's briefly review and discuss the mass spectrum. Numerous lattice
QCD studies have been conducted on the spectrum, and we have compiled
their results alongside ours in Table \ref{The mass calculated by other lattice QCD collaborations}, which are also presented more intuitively in Fig.\ref{fig:Mass}. Regarding the ground $\frac{3}{2}^{+}$ state, most of
the results are in good agreement with each other. Relatively, the
result of Ref \citep{Alexandrou:2012xk} is lower as 4.6769(46)(30).
However, the result is updated to 4.746(4)(32) when simulations were
performed on configurations with the physical pion mass in Ref \citep{Alexandrou:2017xwd}.
Our results for the $\frac{3}{2}^{+}$ state are consistent with the
other lattice calculations within error bars. For the
p-wave $\frac{3}{2}^{-}$ state, the corresponding lattice calculations
are fewer, and the uncertainties are relatively larger. Our results
align closely with those of the TRJQCD collaboration but are slightly
smaller than the results from the HSC and TWQCD collaborations. Considering
that TRJQCD performed simulations on configurations close to the physical
pion mass, the results we obtained here appear reasonable. After our work, two recent studies have also calculated the masses of the triply charmed baryon \citep{Dhindsa:2024erk,Hu:2024mas}. In summary,
the mass spectrum we have obtained is consistent with other lattice
calculations, indicating the reliability of our outcomes within the
current computational constraints.

\begin{table*}[htp]
\caption{The masses of the ground state $\Omega_{ccc}(1^{4}S_{3/2^{+}})$ and
the orbital excited state $\Omega_{ccc}(1^{2}P_{3/2^{-}})$ calculated
by us on $32^{3}\times64$ (48If) and $48^{3}\times96$ (32I) configurations
are shown alongside the results from other lattice QCD collaborations.
The corresponding number of the flavor ($N_{f}$), lattice spacings
(a), the pion mass ($m_{\pi}$), the actions of the relevant sea ($S_{q}^{sea}$)
and valence charm ($S_{c}^{val}$) quarks are also included for comparison.
The abbreviations HISQ and RHQA stand for highly-improved staggered
quark and relativistic heavy-quark action, respectively.}
\setlength{\tabcolsep}{0.1mm}{ \label{The mass calculated by other lattice QCD collaborations}
\centering %
\begin{ruledtabular}

\begin{tabular}{cccccccccccc}

Collaboration  & $N_{f}$  & $a(fm)$  & $m_{\pi}(GeV)$  & $S_{q}^{sea}$  & $S_{c}^{val}$  & $\Omega_{ccc}(\frac{3}{2}^{+})(GeV)$  & $\Omega_{ccc}(\frac{3}{2}^{-})(GeV)$  &  &  &  & 
\tabularnewline
\hline 
\quad{}{[}Ours{]}48If  & 2+1  & 0.0711(3)  & 0.278  & Domain-wall  & Overlap  & 4.793(21)  & 5.071(27) &  &  &  & \tabularnewline
\quad{}{[}Ours{]}32I  & 2+1  & 0.0828(3)  & 0.3  & Domain-wall  & Overlap  & 4.804(20)  & 5.064(51)  &  &  &  & 
\tabularnewline
{[}Ours{]}continuum & 2+1 &  &  &  &  & 4.799(14) & 5.069(24) &  &  &  & 
\tabularnewline 
Y.Lyu et al.~\citep{Lyu:2021qsh} & 2+1 &  0.0846  & 0.146 & Wilson & RHQA &  4.7956(7) &  &  &  &  & 
\tabularnewline
TRJQCD~\citep{Bahtiyar:2020uuj}  & 2+1  & 0.0907(13)  & 0.156(9)  & Clover  & Clover  & 4.817(12)  & 5.083(67)  &  &  &  & \tabularnewline
C.Alexandrou et al.~\citep{Alexandrou:2017xwd}  & 2  & 0.0938(3)(2)  & 0.130  & Twisted Mass   & OS  & 4.746(4)(32)  & -  &  &  &  & \tabularnewline
TWQCD~\citep{Chen:2017kxr}  & 2+1+1  & 0.063  & 0.280  &  Domain-wall  & Domain-wall  & 4.766(5)(11)  & 5.168(37)(51)  &  &  &  & \tabularnewline
K.U.Can~\citep{Can:2015exa}  & 2+1  & 0.0907(13)  & 0.156(7)(2)  & Wilson  & Clover  & 4.769(6)  & -  &  &  &  & \tabularnewline
Z.S.Brown et al.~\citep{Brown:2014ena}  & 2+1  & 0.085-0.11  & 0.227-0.419  & Domain-wall  & RHQA  & 4.796(8)(18)  & -  &  &  &  & \tabularnewline
ETMC~\citep{Alexandrou:2014sha}  & 2+1+1  & 0.065-0.094  & 0.210-0.430  & Twisted Mass  & Twisted Mass  & 4.734(12)(11)(9)  & -  &  &  &  & \tabularnewline
HSC~\citep{Padmanath:2013zfa}  & 2+1  & 0.0351(2)  & 0.390  & Clover  & Clover  & 4.763(6)  & 5.124(13)  &  &  &  & \tabularnewline
PACS-CS~\citep{PACS-CS:2013vie}  & 2+1  & 0.0899  & 0.135(6)  & Clover  & RHQA  & 4.789(22)  & -  &  &  &  & \tabularnewline
R.A.Briceno et al.~\citep{Briceno:2012wt}  & 2+1+1  & 0.06-0.12  & 0.220-0.310  & HISQ  & RHQA  & 4.761(52)(21)(6)  & -  &  &  &  & \tabularnewline
S.Durr et al.~\citep{Durr:2012dw}  & 2  & 0.0728(5)(19)  & 0.280  & Wilson  & Brillouin  & 4.774(24)  & -  &  &  &  & \tabularnewline
C.Alexandrou et al.~\citep{Alexandrou:2012xk}  & 2  & 0.0561(1)-0.089(1)  & 0.260-0.450  & Twisted Mass  & OS  & 4.6769(46)(30)  & -  &  &  &  & \tabularnewline
\end{tabular}

\end{ruledtabular}}
\end{table*}

\subsection{Mass decomposition}

\label{Mass decomposition}

\begin{figure}[htbp]
\centering    
\subfigure{%
\begin{minipage}[l]{9cm}%
 \centering \includegraphics[width=9cm]{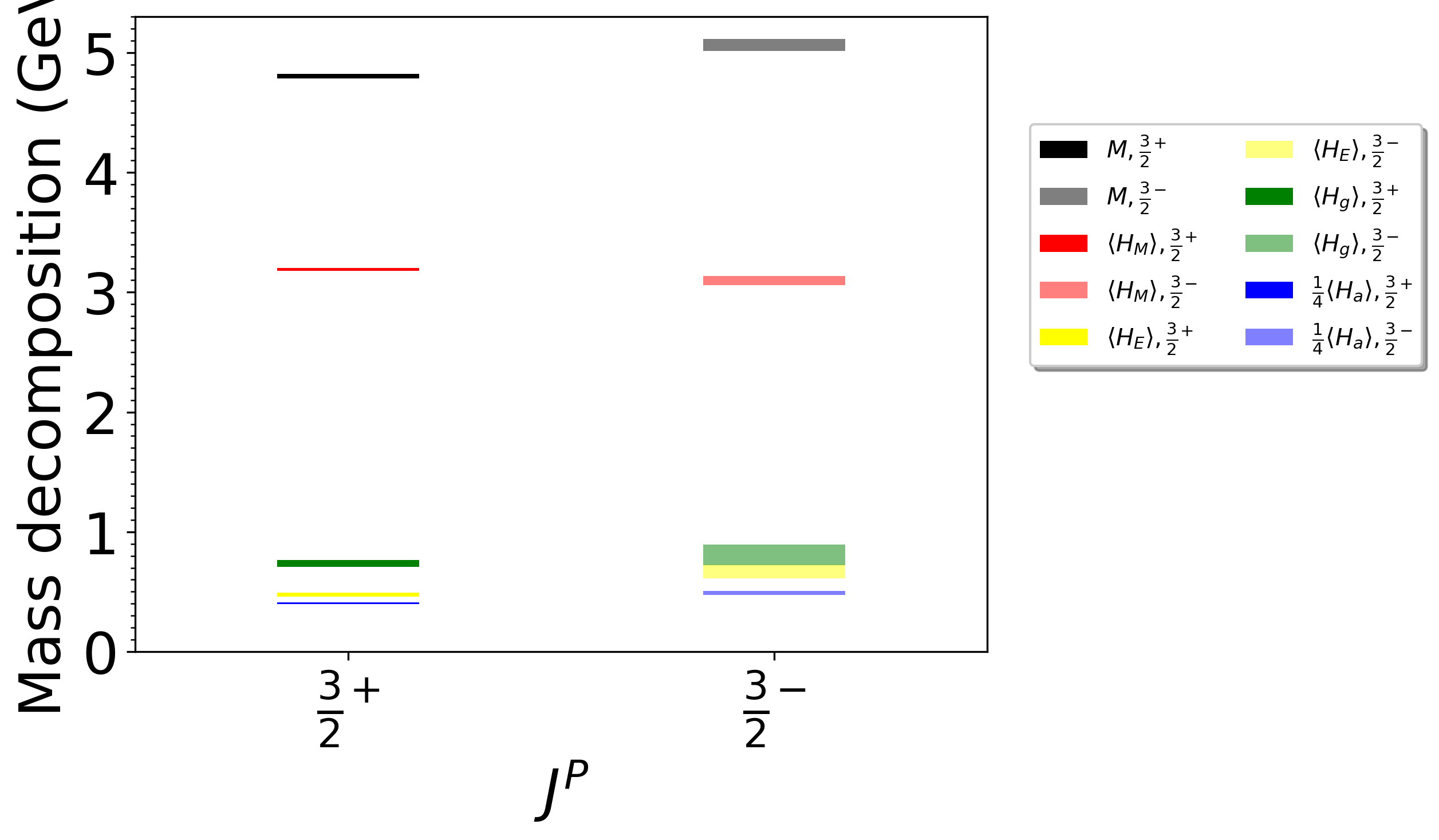} %
\end{minipage}} \subfigure{ %
\begin{minipage}[l]{9cm}%
 \centering   
\includegraphics[width=9cm]{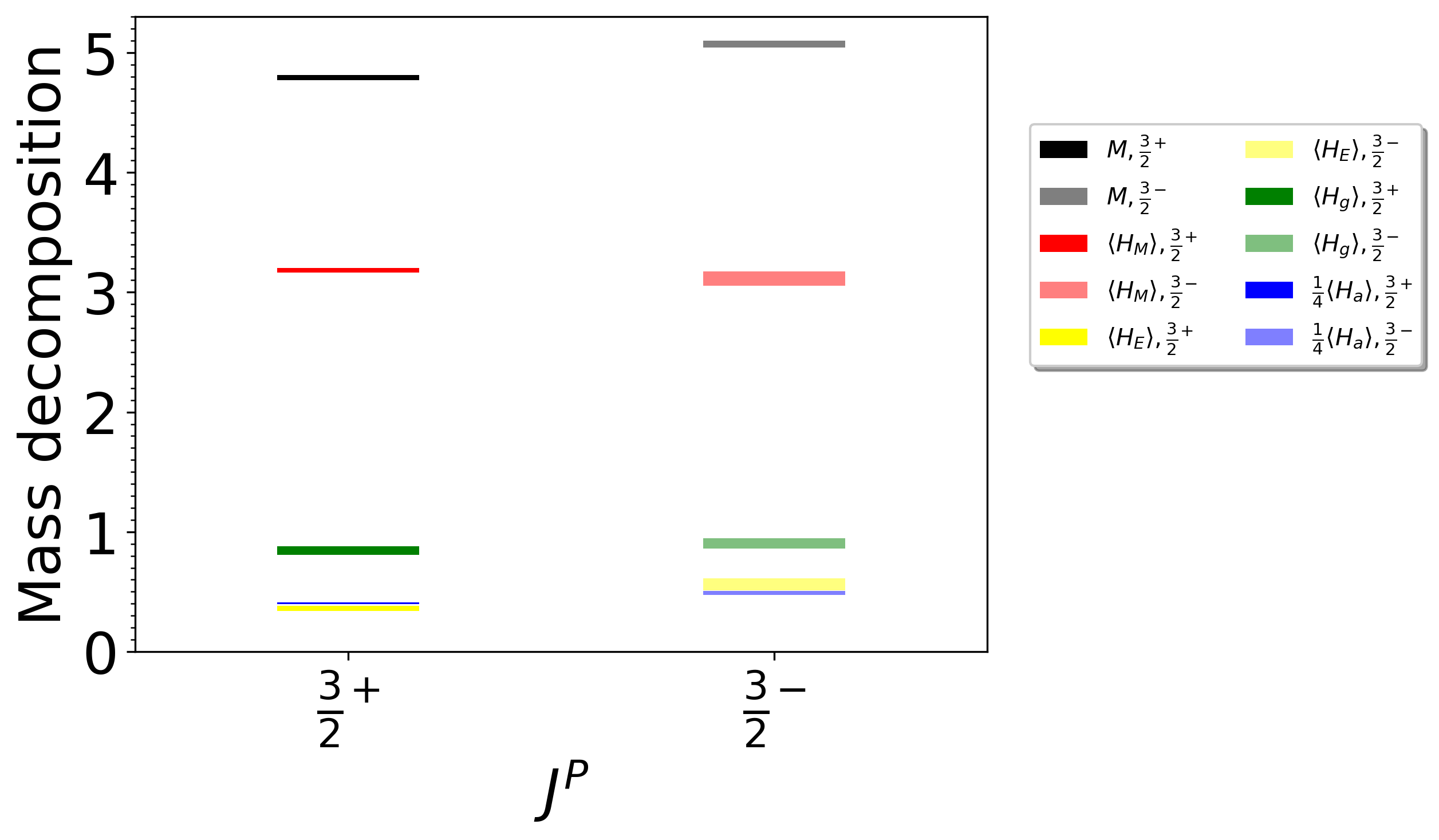} %
\end{minipage}} \caption{Mass decomposition for of $\frac{3}{2}^{+}$ and $\frac{3}{2}^{-}$ triply charmed baryon states on $32^{3}\times64$
(top) and $48^{3}\times96$ (bottom) configurations.}
\label{The result of the mass decomposition of the triply charmed baryon mass} 
\end{figure}

In this work, we have neglected the sea charm quark mass term $\langle H_{M}^{c,\text{sea}}\rangle$
for calculation convenience. This is a reasonable approximation, as
$\langle H_{M}^{c,\text{sea}}\rangle$ is estimated to be less than
100 MeV for both $\frac{3}{2}^{+}$ and $\frac{3}{2}^{-}$ states
on each lattice ensemble, based on the heavy quark expansion
\citep{Shifman:1978zn} as
\begin{equation}
    \left\langle H_{M}^{c,\text{sea}}\right\rangle =\frac{2}{27}\left(\frac{1}{1+\gamma_{m}(\mu)}M-\left\langle H_{M}^{c,v}\right\rangle \right)+O(\alpha_{s}).
\end{equation}
Here, $\gamma_{m}(\mu)\approx\frac{2\alpha(\mu)}{\pi}$ is the quark
anomalous dimension, and $\alpha(\mu=m_{c})\approx0.37$ is taken
from Ref~\citep{Maezawa:2016vgv}. As to the light and strange
sea quarks, we conjecture that it is also safe to neglect their contributions,
based on the observation that their total contribution is less than
40 MeV in charmonium~\citep{Sun:2020pda}. Disconnected
diagrams are also not considered, so their contribution is absorbed
in the QCD anomaly term $H_{a}$ and the gluon energy term $H_{g}$.
Possibly this is the reason why the values of $H_{a}$ and $H_{g}$
we obtained in the triply charmed baryon are slightly larger than
those in charmonium states, as shown in Table \ref{Each part of the triply charmed baryon mass} and more intuitively illustrated in Fig.~\ref{The result of the mass decomposition of the triply charmed baryon mass}. Nevertheless, the total valence charm quark contribution $H_{q}$
remains the main contributor to the triply charmed baryon mass, accounting
for approximately 75\% of the total mass. This is consistent with
the pattern observed in charmonium but differs from that in light
baryons, where the trace anomaly contribution is more significant. 

The quark mass contribution $H_{M}$ in the triply
charmed baryon is about $\frac{3}{2}$ times that in charmonium. This
is consistent with the scenario that a baryon is composed of three
valence quarks, while a meson is composed of two. 
In fact, the quark
condensate contribution is similar to the well-known sigma term in
nucleon physics. Following the definition of the sigma term, we define 
 the renormalized charmness matrix element in the triply charmed baryon as

\begin{equation}
\mathcal{M}_{S}\equiv\langle\Omega_{ccc}(\vec{k}=0)|Z_{S}\bar{\psi^{c}}\psi^{c}|\Omega_{ccc}(\vec{k}=0)\rangle\label{eq:scalar matrix element}
\end{equation}
where $Z_{S}$ is the renormalization constant of the scalar current. As done in Ref.~\cite{Bi:2023pnf, He:2022lse}, we adopt $Z_{S}=[1.009(16),1.008(26)]$ for the ensembles of 32I and 48If, respectively. The charmness matrix element for
the triply charmed baryon $\frac{3}{2}^{+}$ and $\frac{3}{2}^{-}$
states are obtained as shown in Table \ref{Each part of the triply charmed baryon mass},
alongside those of charmonium that we previously calculated. It is
shown that the charmness matrix elements in triply charmed baryons
are greater than that in charmonium. This difference arises primarily
because the scalar current operator couples to the triply charmed
baryon with a factor of 3, whereas the corresponding factor for charmonium
is 2. If the effects of these factors are removed, the relation
$\frac{1}{3}\mathcal{M}_{S}(\Omega)\sim\frac{1}{2}\mathcal{M}_{S}(\bar{c}c)\sim0.9$ is obtained.

Moreover, the mass decomposition provides insights into the nature
of the mass splitting between the s-wave ($\frac{3}{2}^{+}$) and
p-wave ($\frac{3}{2}^{-}$) states. Our calculations indicate a mass splitting of approximately 250 MeV, with the primary contribution stemming from the quark energy term at around 200 MeV.  In contrast, the contributions from the QCD anomaly term $H_{a}$
and the gluon energy term $H_{g}$ are each less than 100 MeV. 
This differs from the scenario in charmonium, where the contributions of $H_{g}$ and $H_{a}$ are comparable to that of $H_{E}$, as evident from the mass decompositions of charmonium listed in Table \ref{Each part of the triply charmed baryon mass}.
For comparison, we have also calculated the rest energy composition
of the ground state ($\frac{3}{2}^{+}$) and orbitally excited state
($\frac{3}{2}^{-}$) triply charmed baryons within the framework of
the constituent quark model. Following Ref~\citep{Liu:2019vtx}, the Hamiltonian for the triply charmed baryon can be expressed as
\begin{equation}
 H=M_{q}+T+\sum_{i,j;i<j}^{3}V_{ij}^{C}+V_{ij}^{G},   
\end{equation}
where $M_{q}$ and $T$ denote the mass and kinetic energy of the
constituent quarks, respectively.  $V^{C}_{ij}$ is the confinement potential defined as $V_{i j}^{C}=\frac{b}{2} r_{i j}$. $V^{G}_{ij}$ is the one-gluon exchange potential, and the explicit expression along with quantities that involved are written as follows, 
\begin{align}
V_{ij}^{G} =& V_{ij}^{coul} + V_{ij}^{sd}, \\ 
V_{ij}^{coul} = & -\frac{2}{3} \frac{\alpha_{i j}}{r_{i j}}, \\
V_{ij}^{sd}= & V_{ij}^{SS}+V_{ij}^{T}+V_{ij}^{LS},\\
V_{ij}^{SS}= & -\frac{2\alpha_{ij}}{3}\left\{ -\frac{\pi}{2}\cdot\frac{\sigma_{ij}^{3}e^{-\sigma_{ij}^{2}r_{ij}^{2}}}{\pi^{3/2}}\cdot\frac{16}{3m_{c}^{2}}\left(\mathbf{S}_{i}\cdot\mathbf{S}_{j}\right)\right\} ,\\
V_{ij}^{T}= & \frac{2\alpha_{ij}}{3}\cdot\frac{1}{m_{c}^{2}r_{ij}^{3}}\left\{ \frac{3\left(\mathbf{S}_{i}\cdot\mathbf{r}_{ij}\right)\left(\mathbf{S}_{j}\cdot\mathbf{r}_{ij}\right)}{r_{ij}^{2}}-\mathbf{S}_{i}\cdot\mathbf{S}_{j}\right\} ,\\
V_{ij}^{LS}= & \frac{\alpha_{\mathrm{SO}}}{\rho^{2}+\lambda^{2}}\cdot\frac{\mathbf{L}\cdot\mathbf{S}}{27m_{c}^{2}}.
\end{align}
Here, $S_i$, S and L are the spin operator of the i-th quark, the total spin of the baryon and the total orbital angular momentum of the baryon, respectively. b, $\alpha_{ij}$, and $\alpha_{\mathrm{SO}}$ denote the strength of confinement potential, strong coupling, and spin-orbit potential. The same parameters are adopted as in Ref~\citep{Liu:2019vtx}.
The corresponding mass decomposition with explicit values of $M_{q},\, T,\, V^{C}, \, V^{coul},\,  V^{SS}, \, V^{T}$ and $V^{SL}$ for both the
$\frac{3}{2}^{+}$ and $\frac{3}{2}^{-}$ states are derived as shown in Table~\ref{Each part of the triply charmed baryon mass calculated from the quark model}.

The constituent quark mass terms for the $\frac{3}{2}^{+}$ and $\frac{3}{2}^{-}$
states are identical, each being three times the constituent quark
mass. Noticed that the constituent quark mass is input as a constant, and the kinetic energy terms are similar for both states. The mass difference between the $\frac{3}{2}^{+}$ and $\frac{3}{2}^{-}$ states primarily arises from the potential energy terms $V^{C}$ and $V^{G}$. The influences of these two potential terms act in opposite directions. The confinement potential
$V^{C}$ has a constructive effect on hadron mass, while the one-gluon
exchange potential $V^{G}$ exerts a destructive influence. The $\frac{3}{2}^{-}$ state has a stronger confinement potential coupled with a weaker one-gluon
exchange potential, leading to a slightly higher mass compared to
the $\frac{3}{2}^{+}$ state, which is consistent with lattice QCD calculations.

It is quite complicated to directly relate lattice
QCD to phenomenological models. However, under certain limits, quantities defined in lattice QCD can reflect phenomenological insights. For example, the gluon trace anomaly term  $\langle H_{a}^{g} \rangle$ for a heavy quark-antiquark pair system is related to the confinement potential by  
\begin{equation}
   \langle H_{a}^{g} \rangle_{\bar{c}c} = A + 2\langle V(r) \rangle, \label{eq:Hag} 
\end{equation}
in the heavy quark limit as demonstrated in Refs.~\citep{Rothe:1995hu,Liu:2021gco}. Here, $A$ is a constant, $V(r) = \sigma r$ is the confinement potential. One may wonder whether this relation still holds for triply charmed baryons. In our calculation, the gluon trace anomaly term $\langle H_{a}^{g}\rangle_{H}  =\langle\frac{\beta}{2g}\int d^{3}\vec{x}G^{2}\rangle_{H}$ can be deduced from 
\begin{equation}
\langle H_{a}^{g}\rangle_{H} =\langle H_{a}\rangle_{H} -\langle H_{a}^{q}\rangle_{H}
\end{equation}
where $\langle H_{a}^{q}\rangle_{H}  =\gamma_{m}\langle H_{m}\rangle_{H}$ denotes the quark trace anomaly term. Here, we focus on the difference of  $\langle H_{a}^{g}\rangle_{H}$ between the $\frac{3}{2}^{+}$ and $\frac{3}{2}^{-}$ states to eliminate the effect of the unknown constant $A$. Adopting $\gamma_{m} = 0.295$ from Ref.~\citep{Hu:2024mas}, we find that the difference of the gluon trace anomaly term is $\Delta\langle H_{a}^{g}\rangle_{\Omega_{ccc}} \equiv\langle H_{a}^{g} \rangle_{\frac{3}{2}^{-}} - \langle H_{a}^{g} \rangle_{\frac{3}{2}^{+}} = 0.384(69)_{\text{32I}}\, \text{GeV  or } 0.369(72)_{\text{48IF}}$ GeV. In comparison, the corresponding difference of the confinement potential in quark model is $\Delta V^{C}(r)\equiv V^{C}(r)_{\frac{3}{2}^{-}}-V^{C}(r)_{\frac{3}{2}^{+}} \simeq 0.16$ GeV.
This approximately satisfies the relation $\Delta\langle H_{a}^{g}\rangle_{\Omega_{ccc}} \simeq 2\Delta \langle V^{C} \rangle$, which is consistent with the charmonium case in the heavy quark limit as denoted by Eq.~(\ref{eq:Hag}). It implies that there may be a connection between the trace anomaly and hadron confinement. Further studies on this issue  would be beneficial.

\begin{table*}
\caption{Mass decomposition of $\frac{3}{2}^{+}$
and $\frac{3}{2}^{-}$  triply charmed baryon states on $32^{3}\times64$
(top) and $48^{3}\times96$ (bottom) configurations, along with the
mass decomposition of charmonium states. The charmness matrix element
$\mathcal{M}_{s}$ are also listed for comparison.}

\label{Each part of the triply charmed baryon mass} \centering %
\begin{ruledtabular}
\begin{tabular}{ccccccc}
$J^{P}$  & $M$(GeV)  & $\langle H_{M}\rangle$(GeV)  & $\langle H_{E}\rangle$(GeV)  & $\langle H_{g}\rangle$(GeV)  & $\frac{1}{4}\langle H_{a}\rangle$(GeV)  & $\mathcal{M}_{S}$\tabularnewline
\hline 
$\frac{3}{2}^{+}$  & 4.804(20)  & 3.192(13)  & 0.474(17)  & 0.735(30)  & 0.403(06)  & 2.742(46)\tabularnewline
$\frac{3}{2}^{-}$  & 5.064(51)  & 3.098(38)  & 0.669(56)  & 0.806(86)  & 0.492(16)  & 2.661(54)\tabularnewline
$J/\psi$ & 3.104(02)  & 2.162(02)  & 0.264(03) & 0.442(02) & 0.2355(07) & 1.857(30)\tabularnewline
$\chi_{c1}$ & 3.434(11)  & 2.101(30)  & 0.335(37) & 0.664(28) & 0.333(08) & 1.805(39)\tabularnewline
\hline 
$\frac{3}{2}^{+}$  & 4.793(21)  & 3.185(19)  & 0.362(22)  & 0.844(37)  & 0.402(07)  & 2.822(76)\tabularnewline
$\frac{3}{2}^{-}$  & 5.071(27)  & 3.114(60)  & 0.564(50)  & 0.904(84)  & 0.489(16)  & 2.759(90)\tabularnewline
$J/\psi$ & 3.100(01)  & 2.139(01) & 0.2116(25) & 0.509(02) & 0.2403(04) & 1.895(50)\tabularnewline
$\chi_{c1}$ & 3.480(18)  & 2.063(38) & 0.387(47) & 0.676(37) & 0.354(11) & 1.828(58)\tabularnewline
\end{tabular}
\end{ruledtabular}
\end{table*}

\begin{table}
\caption{Mass decomposition in quark model. $M$ is the hadron mass of triply
charmed baryon. $M_q$ represents the quark mass term, $T$ denotes the kinetic energy term, $V^C$ stands for the confinement potential, $V^{coul}$ is the Coulomb potential, $V^{SS}$ refers to the spin-spin potential, and $V^{LS}$ is the spin-orbit potential. The tensor potential $V^{T}$ is zero in both of these states and has therefore not been included. The data are given in MeV.}
\label{Each part of the triply charmed baryon mass calculated from the quark model}
\centering %
\begin{ruledtabular}

\begin{tabular}{cccccccc}
$J^{P}$ & $M$ & $M_{q}$& $T$ & $V^{C}$& $V^{coul}$ & $V^{SS}$ & $V^{LS}$\tabularnewline
\hline 
$\frac{3}{2}^{+}$ & 4828 & 4450 & 533 & 471 & -647 & 21 & 0 \tabularnewline
$\frac{3}{2}^{-}$ & 5162 & 4450 & 538 & 630 & -466 & 4 & 7
\tabularnewline
\end{tabular}
\end{ruledtabular}
\end{table}

\section{Summary}\label{Summary}
In this work, we have studied the mass decomposition
of triply charmed baryons within the framework of lattice QCD. The
two lowest triply charmed baryons have been calculated on two lattice ensembles.
An appropriate mass spectrum is obtained, displaying consistency with
the other calculations. The mass decompositions of both $\frac{3}{2}^{+}$
and $\frac{3}{2}^{-}$ triply charmed baryon states on two lattices
are gained. It is found that the total valence charm quark contribution
$H_{q}$ dominate in triply charmed baryon. It is consistent with the pattern observed in charmonium but differs from that displayed in nucleon,
where the trace anomaly contribution plays a major role.

We also calculate the mass decomposition of these two states in the constituent
quark model. It is challenging to directly correlate the mass components decomposed on lattice QCD with those in quark model. Nevertheless, an analysis of the gluon trace anomaly in triply charmed baryons exhibits an analogous relation with the confinement potential to that of heavy quark-antiquark system. This implies a possible connection between the trace anomaly and the confinement potential.

\section*{Acknowledgement}
The authors L. C. Gui, W. Sun and J. Liang, as memebers of the $\chi$QCD corllaboration, thank the RBC collaboration for providing us their DWF gauge configurations. This work is supported by the Natural Science Foundation of China under grant No.12175036, No.11935017, No.12175073, No.12205311, No.12222503. J. Liang is also supported by  Guangdong Major Project of Basic and Applied Basic Research under grant No.2020B0301030008. L. C. Gui is also supported by the Hunan Provincial Natural Science Foundation No.2023JJ30380 and Hunan Provincial Department of Education under grant No.20A310. W. Qin is supported by the Hunan Provincial Natural Science Foundation No.2024JJ6300 and the Scientific Research Fund of Hunan Provincial Education Department (No.22B0044). We thank Hui-hua Zhong for useful discussion. The computations were performed on the Xiangjiang-1 cluster at Hunan Normal University (Changsha) and the Southern Nuclear Science Computing Center (SNSC) and the HPC clusters at Institute of High Energy Physics (Beijing) and China Spallation Neutron Source (Dongguan) and the ORISE Supercomputer.

\bibliography{ref}

\end{document}